\documentclass[...,alpha-refs]{wiley-article}
\usepackage{siunitx}
\usepackage{hyperref} 
\usepackage{graphicx}
\usepackage{adjustbox} % For adjusting table width
\usepackage{array}     % For column definitions
\usepackage{float}
\usepackage{longtable}
\usepackage{orcidlink}
\papertype{Application}

\title{acoupi: An Open-Source Python Framework for Deploying Bioacoustic AI Models on Edge Devices}

\author[1,2\authfn{1}]{Aude Vuilliomenet}
\author[2\authfn{1}]{Santiago Martínez Balvanera}
\author[3]{Oisin Mac Aodha}
\author[2]{Kate E. Jones}
\author[1]{Duncan Wilson}

\contrib[\authfn{1}]{Equally contributing authors.}

\affil[1]{The Bartlett Centre for Advanced Spatial Analysis, Faculty of the Build Environment, University College London, London, W1T 4TJ, United Kingdom}
\affil[2]{Centre for Biodiversity and Environment Research, Department of Genetics, Evolution and Environment, University College London, London, WC1E 6BT, United Kingdom}
\affil[3]{School of Informatics, University of Edinburgh, Edinburgh, United Kingdom}

\corraddress{Aude Vuilliomenet}
\corremail{aude.vuilliomenet.18@ucl.ac.uk}

\fundinginfo{EPSRC, Grant/Award Number: EP/R513143/1 and EP/T517793/1; CONACYT, Grant/Award Number 2020-000017-02EXTF-00334}
\runningauthor{Vuilliomenet et al.}

\begin{document}

\begin{frontmatter}
    \maketitle
    \begin{abstract} \label{sec:abstract}
\textbf{Abstract}
\begin{enumerate}
    \item Passive acoustic monitoring (PAM) coupled with artificial intelligence (AI) is becoming an essential tool for biodiversity monitoring. 
    Traditional PAM systems require manual data offloading and impose substantial demands on data storage and computing infrastructure. 
    The combination of on-device AI-based processing and network connectivity enables to analyse data locally and transmit only relevant information, greatly reducing the volume of data requiring storage. 
    However, programming these devices for robust operation is challenging, requiring expertise in embedded systems and software engineering. 
    Despite the increase in AI-based models for bioacoustics, their full potential remains unrealized without accessible tools to deploy them on custom hardware and tailor device behaviour to specific monitoring goals.
    \item To address this challenge, we develop \texttt{acoupi}, an open-source Python framework that simplifies the creation and deployment of smart bioacoustic devices. \texttt{acoupi} integrates audio recording, AI-based data processing, data management, and real-time wireless messaging into a unified and configurable framework. By modularising key elements of the bioacoustic monitoring workflow, \texttt{acoupi} allows users to easily customise, extend, or select specific components to fit their unique monitoring needs.
    \item We demonstrate the flexibility of \texttt{acoupi} by integrating two bioacoustic classifiers: BirdNET, for the classification of bird species, and BatDetect2, for the classification of UK bat species. We test the reliability of \texttt{acoupi} over a month-long deployment of two \texttt{acoupi}-powered devices in a UK urban park.
    \item \texttt{acoupi} can be deployed on low-cost hardware such as the Raspberry Pi and can be customised for various applications. 
    \texttt{acoupi} standardised framework and simplified tools facilitate the adoption of AI-powered PAM systems for researchers and conservationists.
    \texttt{acoupi} is on GitHub at \url{https://github.com/acoupi/acoupi}.
\end{enumerate}

% Please include a maximum of eight keywords
\keywords{artificial intelligence, biodiversity monitoring, edge computing, passive acoustic monitoring, Raspberry Pi, software}
\end{abstract}
\end{frontmatter}

\section{Introduction} \label{sec:intro}

With the pressing need for biodiversity conservation, recovery, and management \citep{ipbes_global_2019}, it is essential to develop techniques to scale efforts efficiently \citep{besson_towards_2022}. The Kunming-Montreal Global Biodiversity Framework at the 2022 Convention on Biological Diversity Conference (COP15) instituted governments to conserve, manage, and recover natural ecosystems with the goal of protecting 30\% of global land and oceans by 2030 \citep{assembly_transforming_2015, cbd_kunming-montreal_2022}.
This created incentives and obligations to monitor, measure, and track the state of biodiversity, accentuating the need for scalable, affordable, and accessible tools that provide accurate, comprehensive, and transparent biodiversity observation data \citep{stephenson_measuring_2022}. 

Passive acoustic monitoring (PAM) has emerged as a key approach for conducting biodiversity assessments and generating broader ecosystem analyses \citep{browning_passive_2017, gibb_emerging_2019}, while simultaneously providing tools to advance fundamental ecological understanding and conservation science \citep{ross_passive_2023}. The decreasing cost and miniaturisation of audio recording devices, such as the open-source AudioMoth \citep{hill_audiomoth_2019}, have significantly expanded the capacity for deploying extensive acoustic monitoring networks \citep{sethi_safe_2020}. Moreover, the development of AI techniques, including machine learning (ML) and deep learning (DL), to automate the detection and classification of key acoustic signals \citep{kahl_birdnet_2021, mac_aodha_towards_2022} enables researchers to obtain insights at the community and species level from diverse data sets \citep{sethi_large-scale_2024}. This has facilitated long-term acoustic studies across scales, from local to continental \citep{roe_australian_2021}, and in diverse environments, from urban \citep{fairbrass_citynet-deep_2019} to remote and challenging locations \citep{ross_passive_2023}, encompassing terrestrial \citep{sugai_terrestrial_2019} and marine species \citep{mooney_listening_2020}.

\begin{figure}[H]
    \includegraphics[width=1.0\textwidth]{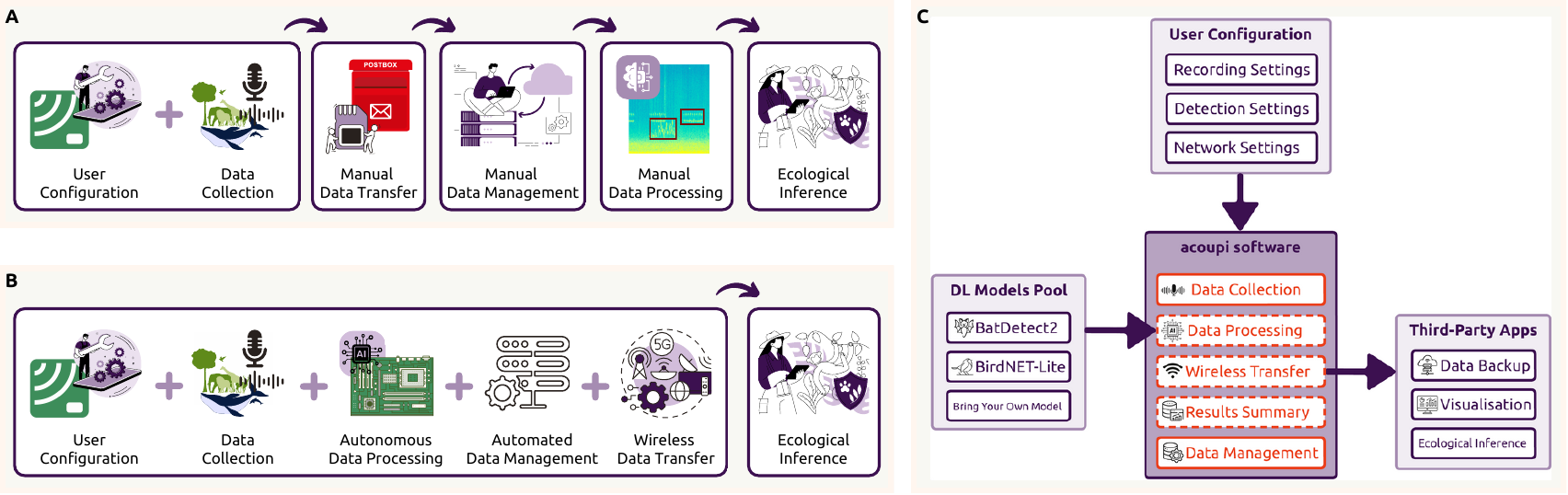}
    \centering
    \caption{\textbf{Overview of acoupi} 
    (A) Traditional PAM workflows consist of fragmented steps requiring frequent intervention, limiting scalability.
    These steps include device deployment, data retrieval and transfer to a central location, data management, data analysis to extract acoustic events, and finally, ecological inference.
    (B) \texttt{acoupi} integrates this workflow into a single device that supports on-board AI-based processing and wireless data transfer, reducing interventions and accelerating data turnaround.
    (C) \texttt{acoupi} adopts a plug-and-play approach, allowing users to configure workflows to their needs.
    Users can specify configuration parameters, select an AI-based model, and set up wireless network endpoints for integration with third-party applications.
    \texttt{acoupi} coordinates essential tasks (orange) like data collection and management, as well as optional modules (dotted) for data processing, transfer and reporting.}
    \label{fig:acoupi_bioacoustic_overview}
\end{figure}

The deployment of PAM systems, however, requires significant human intervention. Frequent site visits are necessary to ensure the correct operation of devices, retrieve audio data, replace storage media, and service batteries \citep{browning_passive_2017}. 
Although stand-alone devices like the AudioMoth \citep{hill_audiomoth_2019} and Solo \citep{whytock_solo_2017} offer configurable recording schedule to adjust sampling effort and optimise resource consumption, monthly visits remain common \citep{karlsson_kinabalu_2021}. 
This creates logistical challenges when maintaining PAM systems, more so in fragmented, large, or remote areas. 
Furthermore, manual retrieval of SD cards and their transport to a central location for analysis \citep{roe_australian_2021, karlsson_kinabalu_2021} introduce risks of data loss and corruption (Fig. \ref{fig:acoupi_bioacoustic_overview}A). 
The inherent physical and spatial separation between data collection and processing introduces significant delays in inferring ecological insights and hampers the timely detection of time-sensitive events, such as illegal hunting activity.

Modern networking technologies can significantly accelerate data transfer from field deployments. The combination of cellular, Wi-Fi, or Long-Range Wide-Area Network (LoRaWAN) communication with continuous power sources like solar panels allows longer deployments without intervention \citep{callebaut_art_2021}. 
Examples include large-scale wildlife monitoring with cellular networks in Borneo \citep{sethi_safe_2020} and Norway \citep{bick_national-scale_2024}, and Wi-Fi network to monitor dolphins in the Mediterranean \citep{brunoldi_permanent_2016}.
However, transferring large audio files, particularly high sample-rate recordings of ultrasonic vocalisations of bats \citep{jones_bat_2007} or rats \citep{coffey_deepsqueak_2019}, can be challenging due to fluctuating cellular data speeds in areas with suboptimal coverage or the inherent bandwidth limitations of LoRaWAN \citep{adelantado_understanding_2017}. 
Critically, storing and processing the transferred recordings is challenging, as large-scale deployments can generate tens of millions of hours of audio, resulting in significant storage and management costs \citep{sethi_robust_2018, roe_australian_2021}. 
Post-deployment processing with AI-based models requires specialised infrastructure and substantial computing power \citep{sethi_safe_2020, stowell_computational_2022}, potentially limiting the accessibility of acoustic monitoring for teams lacking the resources or expertise.

Edge computing \citep{hua_edge_2023}, which involves executing AI-based models on the data-collection devices, offers a promising solution to the challenges of post-deployment processing. 
This approach reduces computational burdens on centralised infrastructure and enhances system responsiveness \citep{baucas_using_2020}. 
Early examples include monitoring bats \citep{zualkernan_aiot_2021, gallacher_shazam_2021}, birds \citep{mcguire_birdnet-pi_2023, disabato_birdsong_2021}, wolves \citep{stahli_development_2022}, as well as urban noise levels \citep{baucas_edge-based_2024} and beehive health \citep{chen_machine_2024}. 
The hardware used in these projects fall broadly into two categories; microcontrollers units (MCUs) and single-board computers (SBCs). 
MCUs are power efficient but have limited compute, restricting the complexity of embeddable AI-based models \citep{disabato_birdsong_2021}, and require proficiency in low-level programming languages to customise.
In contrast, SBCs such as the beginner-friendly and popular Raspberry Pi (RPi) are versatile \citep{jolles_broad-scale_2021} and support various peripherals and sensors.
Additionally, they allow the use of high-level programming languages like Python, a tool increasingly common in ecological research \citep{ulloa_scikit-maad_2021, lapp_opensoundscape_2023, martinez_balvanera_whombat_2024}. 
The BirdNET-Pi project \citep{mcguire_birdnet-pi_2023} illustrates the popularity of setting up RPi-based stations for real-time bird monitoring using the DL model, BirdNET \citep{kahl_birdnet_2021}. 
However, existing edge computing solutions for biodiversity monitoring often use a rigid software architecture that is tightly coupled to specific hardware and selected AI-based model, limiting adaptability.
Developing software to coordinate concurrent recording, processing, and transmission of data presents a technical challenge.
Despite the increasing development of AI-based techniques in bioacoustics \citep{hochst_birdedge_2022}, their potential for biodiversity monitoring remains underused without accessible tools for embedding these models within edge devices while meeting specific project requirements \citep{napier_advancements_2024}.

Here, we develop \texttt{acoupi}, an open-source Python framework that simplifies the creation, configuration, and deployment of devices with on-device processing and network connectivity.
Built with a modular approach, \texttt{acoupi} enables users to configure the entire bioacoustic workflow, from audio data collection and on-device AI processing to data management and wireless transfer (Fig. \ref{fig:acoupi_bioacoustic_overview}B).
Key features include simplified integration of custom AI-based models,  fine-tuning of device behaviour through configuration settings, and reliable deployment on a range of compatible Linux-based SBC devices (Fig. \ref{fig:acoupi_bioacoustic_overview}C).
To demonstrate \texttt{acoupi} capabilities, we integrate two bioacoustic DL classifiers: BirdNET \citep{kahl_birdnet_2021} for avian species and BatDetect2 \citep{mac_aodha_towards_2022} for UK bat species, and evaluate their performance following a month deployment. 
Finally, we discuss the main limitations and highlight key considerations for the effective use of \texttt{acoupi}.

\section{Software Overview} 
\label{sec:software-overview}

The \texttt{acoupi} software is structured in two main parts: a \textit{framework} that provides tools for building programmes and an \textit{application} that manages the configuration and execution of these programmes on edge devices. 
Central to \texttt{acoupi} is the concept of a \textit{``programme''}, defined as a collection of tasks executed by the device. 
Each task represents an independent unit of work, for example data collection, processing, management, or transfer, which together define the behaviour of the bioacoustic sensor (Fig. \ref{fig:acoupi_program}).
The \textit{framework} provides a flexible yet structured and standardised approach for defining programmes.
The \textit{application} ensures the harmonious and fault-tolerant execution of a programme. 
Moreover, it allows users to customise programmes parameters via a simple command-line interface (CLI), facilitating a ``no-code'' approach.

\subsection{acoupi Framework}
The \texttt{acoupi} framework is designed to simplify the creation of customised programmes. 
While customisability is a primary objective, a key secondary goal is programme standardisation. 
This is achieved by enforcing that programme inputs, behaviours, and outputs are consistent across all programmes. 
This standardisation offers several advantages: it guarantees compatibility of user customisations within the framework, facilitates collaborations and sharing among users, and eases integration with other devices and third-party services. 
Standardisation also establishes a common language for understanding and discussing programme design. 
To achieve these objectives, the \texttt{acoupi} framework provides a suite of tools organised into four layers of increasing flexibility. 
This section introduces each layer, highlighting their role in programme creation and customisation. 

\begin{figure}[H]
\includegraphics[width=1\textwidth]{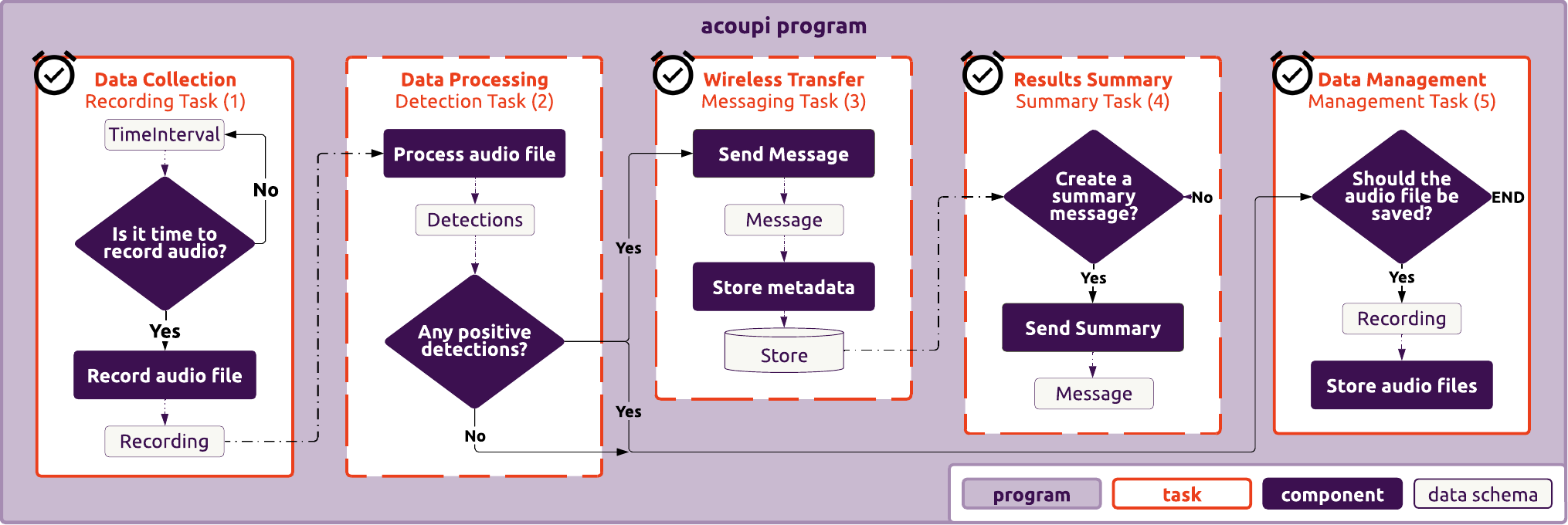}
\centering
\caption{
\textbf{Example of a simplified \texttt{acoupi} programme.} 
This programme (mauve) implements five tasks: (1) recording, (2) detection, (3) messaging, (4) summary, and (5) management. 
Each task (orange) follows a standardised workflow of individual components (dark purple), involving actions (rectangles) and decisions (rhombuses). Tasks can be scheduled to occur at regular time intervals (indicated by a clock), or triggered by other tasks (without a clock).  
Users can exchange components to modify device behaviour, customising how actions are performed and decisions are made without altering the overall workflow.
Component behaviour can be fine-tuned through user-provided configuration parameters.
Standardised data objects (light grey) flow between components, ensuring consistency across the workflow.
}
\label{fig:acoupi_program}
\end{figure}

The \textbf{programme} layer provides \textit{programme templates} that require minimal modification to create fully functional programmes. 
Each template consists of a set of operations that can be extended or adapted (Fig. \ref{fig:acoupi_program}). 
For example, the \textit{BasicProgram} template includes the minimum operations for a valid \texttt{acoupi} programme, corresponding to recording and saving of audio files.
In contrast, the \textit{DetectionProgram} template offers the full set of \texttt{acoupi} features, providing tasks to record audio, process recordings using an AI-based model, transfer results wirelessly to a remote server, manage the saving and deletion of audio files and metadata, and create periodic summaries of the system.

The \textbf{tasks} layer provides additional control to the pre-defined programme templates. 
In \texttt{acoupi}, a task can be any user-defined Python function, yet to facilitate standardisation and streamline development, the framework offers six \textit{task templates} covering common operations. 
These include: 
(1) \textit{recording} to collect audio data following a user-defined schedule, 
(2) \textit{detection} to process captured recordings, 
(3) \textit{messaging} to transmit detections to a remote server, 
(4) \textit{summary} to generate periodic analytical reports, 
(5) \textit{management} to organise data storage and handle audio files, and 
(6) \textit{heartbeat} to monitor system health. 
Tasks can either be scheduled, for example, to record audio at regular intervals or triggered by other tasks, such as using a DL model to process a recording file as soon as a recording finishes. The tasks built with these templates follow a standardised sequence of actions and decisions, which are implemented by the user-provided components (Fig. \ref{fig:acoupi_program}). 

The \textbf{components} are reusable and perform one specific operation based on the configuration input of an \texttt{acoupi} programme. 
All components adhere to a set of definitions, called component types, which specify the functionality of a component, its inputs, and outputs. 
For instance, the \textit{Storage} component provides users with the ability to interact with a local database to store metadata of captured recordings, or to store the detections and classifications outputs after a recording has been processed.
Furthermore, the \textit{RecordingSavingFilter} component allows audio files to be saved based on criteria such as recording time, detection confidence score or specific classification tags.

The \textbf{data schema} layer defines a set of ``data objects'' made of specific data types (e.g., string, float, path). These data objects standardise data exchange between components. They are packets of information that are generated during the execution of a programme. Examples of data objects are \texttt{Recording}, \texttt{Detection}, and \texttt{Message}. These objects help to ensure consistency and compatibility throughout the programme flow.

A comprehensive overview of the programmes, tasks, components and data schemas, along with step-by-step instructions for modifying and creating new programmes, can be found in the documentation under the \href{https://acoupi.github.io/acoupi/howtoguide/programs/}{``How-To Guides: Create a custom programme''}.

\subsection{acoupi Application}

\begin{figure}[H]
\includegraphics[width=1.0\textwidth]{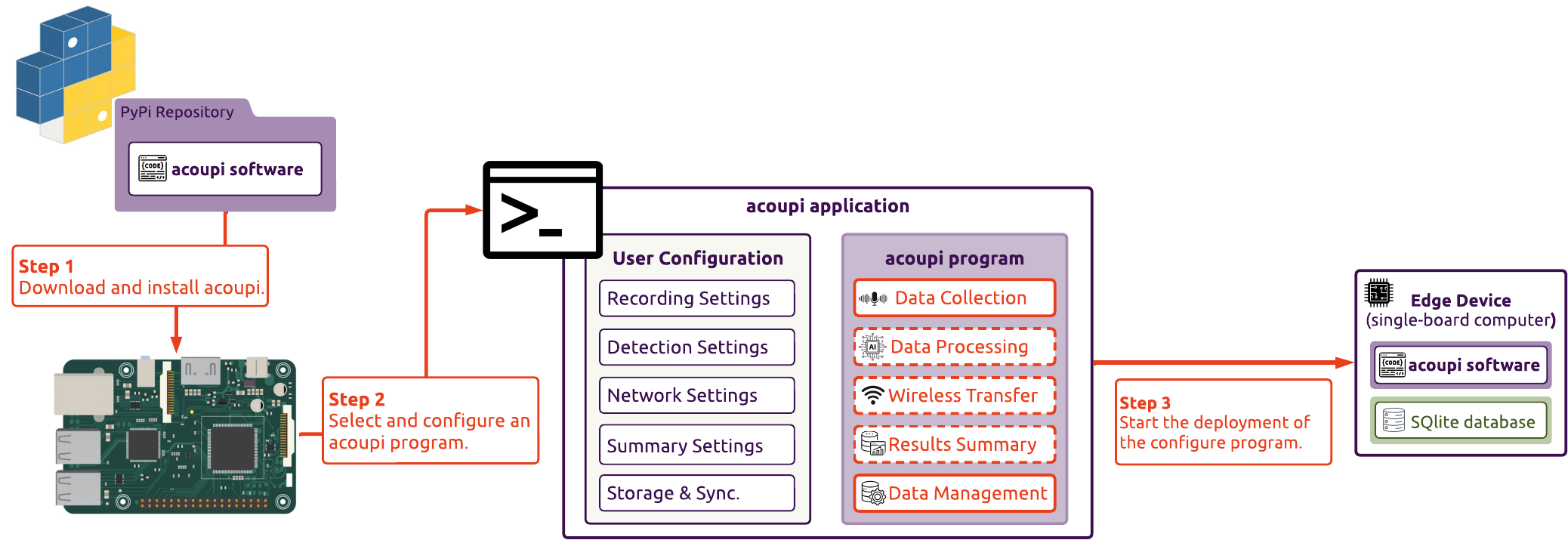}
\centering
\caption{\textbf{Overview of the steps to install, configure, and start an \texttt{acoupi} application.}. Step (1) downloads and installs the \texttt{acoupi} software from the PyPi repository on a single-board computer. In step (2), users select and input the configuration parameters of an \texttt{acoupi} programme via a command-line interface. The configuration parameters define how the recording, processing, messaging, summary, and management tasks are executed. In step (3), users start a deployment running the configured \texttt{acoupi} programme.}
\label{fig:acoupi_application}
\end{figure}

The \texttt{acoupi} application enables the execution of a pre-built programme on a chosen edge device (Fig. \ref{fig:acoupi_application}). 
The application provides a command-line interface (CLI) with simple commands to manage and deploy programmes.
The command \textit{``acoupi setup''} guides users through a configuration wizard, allowing them to select a programme and configure its parameters.
The validity of the configurations can be checked and modified using the command \textit{``acoupi config''}.
Once a programme is configured, users can initiate deployment with the command \textit{``acoupi deployment start''}.
The application performs pre-deployment health checks to verify the programme configuration and system setup, identifying potential issues such as connectivity problems or microphone malfunctions.
Finally, the command \textit{``acoupi deployment stop''} shuts down the system and records the start and end times of a deployment.

The \texttt{acoupi} application ensures the timely execution of programme tasks. 
Maintaining reliable operation on edge devices can be difficult due to computational resource limitations, network instability, and power fluctuations.
To address this, \texttt{acoupi} leverages Celery \citep{solem_celery_2025}, a robust and widely-used task management tool. 
Celery helps coordinate and schedule tasks, automatically retries failed tasks, and runs multiple tasks simultaneously whenever possible.
Furthermore, \texttt{acoupi} incorporates mechanisms for automatic recovery after power failures. A comprehensive log of device activities is maintained aiding in identifying failures and preventing data loss.

To optimise storage usage during deployments, \texttt{acoupi} does not store recordings by default.
Instead, recordings are temporarily held in the working memory for processing.
Depending on the programme's logic and configuration, recordings may be selectively saved, such as when an AI-based model identifies vocalisations of target species.
Both the chosen programme and its configurations are stored, facilitating the reproducibility of deployments by sharing configuration files. 
Additionally, \texttt{acoupi} stores lightweight SQLite databases containing essential recording metadata and automated detections.
This design helps mitigate the risk of premature termination of the deployment due to storage depletion while maintaining crucial metadata for subsequent analysis and reproducibility.

\subsection{Requirements}

To run \texttt{acoupi}, a single-board computer (SBC) running a Linux operating system (OS) is required.
\texttt{acoupi} has been extensively tested on a Raspberry Pi 4 Model B running the 64-bit RPi OS; however, devices with similar specifications should be compatible.
RPi boards are recommended for new users due to their beginner-friendliness and extensive documentation \citep{jolles_broad-scale_2021}.

In addition to a SBC, a microphone and a microSD card are required.
The microphone should be selected according to the desired sampling rate, which depends on the target species' vocalisation frequencies.
To ensure adequate capture, the sampling rate should be greater than twice the highest frequency of the target vocalisation.
A microSD card with a minimum capacity of 32GB is recommended.
Users should select a larger capacity microSD card or consider an external hard drive, based on the volume of audio files they wish to archive for offline analysis post-deployment. 

A detailed step-by-step installation is available at \url{acoupi.github.io/acoupi/#installation}. 
\section{Pre-Built programmes}
\label{sec:prebuilt-programmes}

We develop two ready-to-use programmes: \textit{acoupi\_birdnet} and \textit{acoupi\_batdetect2}.
These programmes offer out-of-the-box functionality that can be customised through configuration adjustments without requiring any coding.
Both programmes are built upon the \texttt{DetectionTemplate} described in the previous section and thus inherit a common structure while incorporating distinct AI-based models and default configurations.

These programmes leverage two established DL models for the acoustic detection of birds and bats.
\textit{acoupi\_birdnet} employs the BirdNET model version 2.4 \citep{kahl_birdnet_2021}, capable of detecting approximately 6,400 bird species globally, along with other relevant acoustic events such as frog calls, domestic animals, fireworks, and engine noise.
\textit{acoupi\_batdetect2} utilises the BatDetect2 model \citep{mac_aodha_towards_2022}, designed to detect echolocation calls from 17 bat species commonly found in the UK.
BirdNET and BatDetect2 models were trained using recordings made at 48kHz and 256kHz, respectively, thus using the same or similar sampling rate when using these models is recommended for better performance.
These models showed good performance within the scope of their original evaluations \citep{perez-granados_birdnet_2023}. 
As with all DL models, there is the potential for misidentification or missed detections, particularly in environments dissimilar to their training data \citep{van_merrienboer_birds_2024}.
Thorough evaluation of model performance within the specific deployment context is strongly recommended \citep{wood_real-time_2024}.

Both \textit{acoupi\_birdnet} and \textit{acoupi\_batdetect2} feature automated and scheduled recording, processing with their respective DL models, and transmission of detections to a remote server.
\textit{acoupi\_birdnet} captures nine seconds every ten seconds between 03:00 and 23:00, while \textit{acoupi\_batdetect2} records three seconds of audio at the same frequency between 19:00 and 07:00.
Detections exceeding a predefined confidence threshold are transmitted to a remote server every 30 seconds.
A heartbeat signal is transmitted every 30 minutes to monitor device health, regardless of recording activity or detection events.
The programmes allow for optional storage of recordings with confident detections, facilitating post-deployment validation.
Crucially, all operational parameters, including recording schedules, durations, frequencies, and messaging intervals, are fully configurable.
It is essential to carefully consider the monitoring goals and adjust these settings accordingly \citep{teixeira_effective_2024}.
\section{Software Testing}
\label{sec:software-testing}

To test the reliability of \texttt{acoupi}, we configured and deployed both the \textit{acoupi\_birdnet} and \textit{acoupi\_batdetect2} programmes on two separate RPi 4B devices.
The RPis were deployed in the People and Nature Garden Lab on the roof of UCL One Pool Street building within the Queen Elizabeth Olympic Park in London, UK (51° 32' 18.8'' N, 0° 0' 32.36'' W).
This location was selected for initial software testing due to convenient access to power and Wi-Fi network, acknowledging that such conditions may not fully represent the challenges of field deployments. 
Most of the configuration parameters for the \textit{acoupi\_birdnet} and \textit{acoupi\_batdetect2} programmes followed the default configuration (Appendix \ref{appendix:configuration_table}).
The devices were deployed for 30 days between October and November 2024.

\begin{table}[H]
\centering
\renewcommand{\arraystretch}{2} % Adjust row height for better spacing
\adjustbox{width=\textwidth}{%
\fontsize{16}{18}\selectfont

\begin{tabular}{|>{\raggedright\arraybackslash}p{6cm}|>{\raggedright\arraybackslash}p{4.5cm}|>{\centering\arraybackslash}p{4cm}|>{\centering\arraybackslash}p{4cm}|>{\centering\arraybackslash}p{4cm}|>{\centering\arraybackslash}p{4cm}|} \hline
& &
\multicolumn{2}{c|}{\textbf{Processing Failures}} & 
\multicolumn{2}{c|}{\textbf{Messages Sent}} \\ \hline
\textbf{Programme} & 
\textbf{\# Recordings} & 
\textbf{Count} & 
\textbf{\% Total} & 
\textbf{Count} & 
\textbf{\% Total} \\ \hline
\textit{acoupi\_ birdnet} & 129,939 & 4 & 0.003\% & 8716 & 100\%  \\ \hline
\textit{acoupi\_ batdetect2} & 65,711  & 1,203 & 1.83\% & 868 & 100\% \\ \hline
\end{tabular}%
} 
\caption{\textbf{Deployment metrics of \textit{acoupi\_birdnet} and \textit{acoupi\_batdetect2}.} Key metrics from a month deployment, including the total number of recordings, the count (and percentage) of unprocessed recordings, and the count (and percentage) of messages successfully sent to the remote server.}
\label{tab:software_testing_summary_deployment}
\end{table}

To evaluate the reliability of the software, we examined key metrics, including recording consistency, processing success, and message delivery (Table \ref{tab:software_testing_summary_deployment}).
Both programmes successfully recorded at every scheduled interval; however, minor
imprecisions in the scheduler resulted in an average recording frequency of approximately 10.005 seconds (Appendix \ref{appendix:deployment_results}). 
Most of the recordings were successfully processed by the integrated DL models.
The \textit{acoupi\_birdnet} processed all but four recordings, while the \textit{acoupi\_batdetect2} exhibited a slightly higher failure rate, with 1.83\% of recordings not being processed. 
Both programmes reliably delivered all generated messages, indicating stable message delivery under good network conditions.
Despite these positive results, both deployments encountered premature termination.
The device running \textit{acoupi\_birdnet} was likely dislodged by strong winds, resulting in power loss.
The \textit{acoupi\_batdetect2} programme encountered a software issue that accounted for the 1,203 unprocessed recordings. 
This issue has since been resolved. For a summary of the detections made by the two \texttt{acoupi} programmes, see Appendix \ref{appendix:detection_results}.

\section{Limitations and Future Directions}
\label{sec:limitations}

The present work showed how \texttt{acoupi} can be used to embed two bioacoustic DL models, BirdNET and BatDetect2. 
Although BirdNET covers a wide range of avian species \citep{kahl_birdnet_2021}, and BatDetect2 targets all bat species found in the UK \citep{mac_aodha_towards_2022}, it will be necessary to integrate additional AI-based models to accommodate a greater diversity of species and applications.
Future updates could include any other AI-based models, such as the animal detection toolbox, Tadarida \citep{bas_tadarida_2017} and the frog detector, RIBBIT \citep{lapp_automated_2021}. 
However, model integration requires considering their size and complexity, as these factors affect processing speed and power consumption on edge devices \citep{desislavov_trends_2023}. 
If processing times exceed the recording interval, processing backlogs and system overloads can occur. 
Optimisation techniques like quantisation \citep{krishnamoorthi_quantizing_2018, novac_quantization_2021}, pruning \citep{hoeer_sparsity_2021}, and knowledge distillation \citep{gou_knowledge_2021} can reduce computational demands, but must be applied carefully to balance model compression and accuracy \citep{desislavov_trends_2023}. 
As models optimised for edge processing emerge \citep{disabato_birdsong_2021, zualkernan_aiot_2021, hochst_birdedge_2022, ghani_global_2023}, \texttt{acoupi} will provide a platform for their integration, making these advances accessible to the community. 

In its month-long deployment at the People and Nature Lab in London, UK, \texttt{acoupi} successfully captured all scheduled recordings, sent all detection messages, and processed the majority of recordings using the embedded DL models. 
Nevertheless, the premature termination of both deployments underscores the need for more extensive field testing.
While \texttt{acoupi} leverages established software tools to ensure reliable operation, empirical evaluations under varying network connectivity and power availability remain essential. 
Additionally, although \texttt{acoupi} is designed to operate on any Linux-compatible SBC, further testing on various SBCs would provide valuable insight into its compatibility and performance across different hardware platforms. 
Future \texttt{acoupi} development will focus on addressing these challenges through iterative improvements informed by field tests. 
The codebase includes automated tests to facilitate modifications as well as a system to distribute updates remotely, thus supporting the long-term maintenance and adaptability of the \texttt{acoupi} software.

In its first version, \texttt{acoupi} is limited to audio recordings as its primary data collection method and relies on Wi-Fi for communication. 
However, RPi-type boards support additional sensors and alternative connectivity such as cellular networks and LoRaWAN \citep{jolles_broad-scale_2021}. 
One promising extension is the integration of multichannel audio recorders \citep{heath_spatial_2024}, allowing on-board localisation algorithms to estimate the position of vocalising animals, a critical step for improving population density estimates \citep{rhinehart_acoustic_2020}. 
Furthermore, additional sensors to capture abiotic data, such as rainfall, wind speed, luminosity, temperature, and humidity, could provide an environmental context for ecological analyses, as these factors influence sound transmission and species detectability \citep{ross_utility_2021, metcalf_good_2023}. 
Low-cost camera modules could also transform \texttt{acoupi} deployments into smart camera trap systems \citep{darras_eyes_2024} combining visual and acoustic monitoring. 
While these extensions await implementation, \texttt{acoupi}'s modular framework provides the foundation for community-driven contribution and on-going development. 

This work highlights the flexibility of \texttt{acoupi}, though its requirements may limit its use in certain monitoring scenarios.
\texttt{acoupi} is designed for SBCs, which consume more power than MCUs, and requires a continuous power supply to ensure uninterrupted operation through solar panels with batteries or a connection to mains power.
A complete deployment necessitates additional components, including a microphone and a protective enclosure, increasing overall cost and complexity.
Whereas the base cost of an \texttt{acoupi}-compatible SBC is comparable to that of an AudioMoth, a full setup may exceed the budget of resource-constrained projects (see \cite{sethi_robust_2018} for the cost breakdown of an analogous system). 
Despite these limitations, \texttt{acoupi} is particularly well-suited for continuous and long-term monitoring of key sites with stable power and network access. 
Future iterations could incorporate power management mechanisms, like intelligent scheduling to optimise detection probability while minimising power use \citep{balle_power_2024, millar_terracorder_2024}. 

The use of bioacoustic monitoring tools such as \texttt{acoupi} raises important ethical questions, particularly regarding privacy \citep{sandbrook_principles_2021}. 
PAM systems lack the ability to differentiate between wildlife vocalisations and human speech, leading to  unintended capture of human conversations where consent is absent. 
Although researchers must adhere to responsible data handling practices like minimal data collection, secure storage, and timely deletion \citep{zook_ten_2017}, data upload to cloud-based storage and network devices remain vulnerable to security breaches and malicious attacks \citep{kim_survey_2022}. 
The on-device processing in \texttt{acoupi} offers a privacy advantage by allowing identification and immediate deletion of audio clips containing human voices \citep{cretois_voice_2022}. 
However, the initial release does not include human speech detection functionality and, therefore, users are strongly encouraged to ensure compliance with privacy regulations. 

Ultimately, \texttt{acoupi} aims to provide a flexible, user-friendly tool for bioacoustic monitoring across diverse monitoring scenarios. 
For example, real-time detections generated by \texttt{acoupi} could be integrated into dashboards to create immediate alerts for time-sensitive events, such as mitigating human-wildlife conflicts, managing invasive species \citep{wood_real-time_2024}, or dimming city lights in response to migratory bird movements \citep{horton_bright_2019}.
Beyond real-time applications, these detections can support ecological research through methods like occupancy modelling \citep{rhinehart_continuous-score_2022} and call density analysis \citep{navine_all_2024}.
The validation of automated detections remains crucial when using AI models in novel environments \citep{perez-granados_birdnet_2023, van_merrienboer_birds_2024}.
\texttt{acoupi} can facilitate this validation by storing recordings that exceed a detection score threshold, or other criteria aligned with specific modelling requirements \citep{navine_all_2024, knight_validation_2020}. 
By establishing standardised concepts, each with specified metadata, \texttt{acoupi} promotes integration and comparability of monitoring surveys, enabling ecologists to share findings, compare results between sites, and ensure reproducibility \citep{besson_towards_2022}.

As a final point, it is important to highlight the open-source nature of \texttt{acoupi}, which allows researchers to adapt and extend \texttt{acoupi}'s customisable software modules to their specific requirements. 
We acknowledge that \texttt{acoupi} pre-configured programmes may not meet the needs of every researcher, and therefore invite and encourage the community to contribute to expand \texttt{acoupi}'s functionalities.
At its heart, \texttt{acoupi} aims to provide greater accessibility to bioacoustic monitoring, enabling real-time and automated monitoring for ecological research and conservation efforts.

\section*{Acknowledgements}
This work was financially supported by URKI EPSRC (EP/R513143/1 and EP/T517793/1) and Consejo Nacional de Ciencia y Tecnologia (CONACYT) (2020-000017-02EXTF-00334). We thank Ella Browning for her feedback on the design of acoupi. 
\section*{Conflicts of Interest}
The authors declare no conflict of interest.
\section*{Authors' Contribution}
Aude Vuilliomenet: conceptualisation (equal); investigation (equal); software (supporting); visualisation (lead); writing – original draft (lead).
Santiago Martinez Balvanera: conceptualisation (equal); investigation (equal); software (lead); visualisation (supporting); writing – original draft (supporting).
Oisin Mac Aodha: supervision (equal); writing – review and editing (equal).
Kate E. Jones: resources (equal); supervision (equal); writing - review and editing (equal)
Duncan Wilson: resources (equal); supervision (equal); writing – review and editing (equal)
\section*{Code and Data Availability Statement}
acoupi is available on PyPi and GitHub at \url{github.com/acoupi/acoupi}. The documentation provides detailed instructions on how to use and contribute to acoupi. It is available in English at \url{acoupi.github.io/acoupi}.
\section*{ORCID}
\textit{Aude Vuilliomenet} \orcidlinkf{0000-0003-1517-9296} \\
\textit{Santiago Martinez} Balvanera \orcidlinkf{0000-0003-1214-1938} \\
\textit{Oisin Mac Aodha} \orcidlinkf{0000-0002-5787-5073} \\
\textit{Kate E. Jones }\orcidlinkf{0000-0001-5231-3293} \\
\textit{Duncan Wilson} \orcidlinkf{0000-0001-6041-8044} \\

\bibliography{references}

\begin{thebibliography}{70}
\expandafter\ifx\csname natexlab\endcsname\relax\def\natexlab#1{#1}\fi
\expandafter\ifx\csname url\endcsname\relax
  \def\url#1{\texttt{#1}}\fi
\expandafter\ifx\csname urlprefix\endcsname\relax\def\urlprefix{URL: }\fi

\bibitem[{Adelantado et~al.(2017)Adelantado, Vilajosana, Tuset-Peiro, Martinez, Melia-Segui and Watteyne}]{adelantado_understanding_2017}
Adelantado, F., Vilajosana, X., Tuset-Peiro, P., Martinez, B., Melia-Segui, J. and Watteyne, T. (2017) Understanding the {Limits} of {LoRaWAN}.
\newblock \textit{IEEE Communications Magazine}, \textbf{55}, 34--40.
\newblock \urlprefix\url{https://ieeexplore.ieee.org/document/8030482}.
\newblock Conference Name: IEEE Communications Magazine.

\bibitem[{Assembly(2015)}]{assembly_transforming_2015}
Assembly, U.~G. (2015) Transforming our world: the 2030 {Agenda} for {Sustainable} {Development}.
\newblock \textit{Seventieth session A/RES/70/1}, United Nations, New York, USA.

\bibitem[{Balle et~al.(2024)Balle, Xu, Darras and Wanger}]{balle_power_2024}
Balle, M., Xu, W., Darras, K.~F. and Wanger, T.~C. (2024) A {Power} {Management} and {Control} {System} for {Environmental} {Monitoring} {Devices}.
\newblock \textit{IEEE Transactions on AgriFood Electronics}, 1--10.
\newblock \urlprefix\url{https://ieeexplore.ieee.org/document/10750140}.
\newblock Conference Name: IEEE Transactions on AgriFood Electronics.

\bibitem[{Bas et~al.(2017)Bas, Bas and Julien}]{bas_tadarida_2017}
Bas, Y., Bas, D. and Julien, J.-F. (2017) Tadarida: {A} {Toolbox} for {Animal} {Detection} on {Acoustic} {Recordings}.
\newblock \textit{Journal of Open Research Software}, \textbf{5}.
\newblock \urlprefix\url{https://openresearchsoftware.metajnl.com/articles/10.5334/jors.154}.

\bibitem[{Baucas and Spachos(2020)}]{baucas_using_2020}
Baucas, M.~J. and Spachos, P. (2020) Using cloud and fog computing for large scale {IoT}-based urban sound classification.
\newblock \textit{Simulation Modelling Practice and Theory}, \textbf{101}, 102013.
\newblock \urlprefix\url{https://www.sciencedirect.com/science/article/pii/S1569190X19301431}.

\bibitem[{Baucas and Spachos(2024)}]{baucas_edge-based_2024}
--- (2024) Edge-{Based} {Data} {Sensing} and {Processing} {Platform} for {Urban} {Noise} {Classification}.
\newblock \textit{IEEE Sensors Letters}, \textbf{8}, 1--4.
\newblock \urlprefix\url{https://ieeexplore.ieee.org/document/10506670/}.

\bibitem[{Besson et~al.(2022)Besson, Alison, Bjerge, Gorochowski, Høye, Jucker, Mann and Clements}]{besson_towards_2022}
Besson, M., Alison, J., Bjerge, K., Gorochowski, T.~E., Høye, T.~T., Jucker, T., Mann, H. M.~R. and Clements, C.~F. (2022) Towards the fully automated monitoring of ecological communities.
\newblock \textit{Ecology Letters}, \textbf{25}, 2753--2775.
\newblock \urlprefix\url{https://onlinelibrary.wiley.com/doi/abs/10.1111/ele.14123}.
\newblock \_eprint: https://onlinelibrary.wiley.com/doi/pdf/10.1111/ele.14123.

\bibitem[{Bick et~al.(2024)Bick, Bakkestuen, Cretois, Hillier, Kålås, Pedersen, Raja, Rosten, Somveille, Stokke, Wiel and Sethi}]{bick_national-scale_2024}
Bick, I.~A., Bakkestuen, V., Cretois, B., Hillier, B., Kålås, J.~A., Pedersen, M., Raja, K., Rosten, C.~M., Somveille, M., Stokke, B.~G., Wiel, J. and Sethi, S.~S. (2024) National-scale acoustic monitoring of avian biodiversity and migration.
\newblock \urlprefix\url{http://biorxiv.org/lookup/doi/10.1101/2024.05.21.595242}.
\newblock 1 citations (Semantic Scholar/DOI) [2024-12-09].

\bibitem[{Browning et~al.(2017)Browning, Gibb, Glover-Kapfer and Jones}]{browning_passive_2017}
Browning, E., Gibb, R., Glover-Kapfer, P. and Jones, K.~E. (2017) Passive acoustic monitoring in ecology and conservation.
\newblock \textit{{WWF} {Conservation} {Technology}}, WWF-UK, Woking, United Kingdom.
\newblock \urlprefix\url{https://www.wwf.org.uk/sites/default/files/2019-04/Acousticmonitoring-WWF-guidelines.pdf}.

\bibitem[{Brunoldi et~al.(2016)Brunoldi, Bozzini, Casale, Corvisiero, Grosso, Magnoli, Alessi, Bianchi, Mandich, Morri, Povero, Wurtz, Melchiorre, Viano, Cappanera, Fanciulli, Bei, Stasi and Taiuti}]{brunoldi_permanent_2016}
Brunoldi, M., Bozzini, G., Casale, A., Corvisiero, P., Grosso, D., Magnoli, N., Alessi, J., Bianchi, C.~N., Mandich, A., Morri, C., Povero, P., Wurtz, M., Melchiorre, C., Viano, G., Cappanera, V., Fanciulli, G., Bei, M., Stasi, N. and Taiuti, M. (2016) A {Permanent} {Automated} {Real}-{Time} {Passive} {Acoustic} {Monitoring} {System} for {Bottlenose} {Dolphin} {Conservation} in the {Mediterranean} {Sea}.
\newblock \textit{PLOS ONE}, \textbf{11}, e0145362.
\newblock \urlprefix\url{https://journals.plos.org/plosone/article?id=10.1371/journal.pone.0145362}.
\newblock Publisher: Public Library of Science.

\bibitem[{Callebaut et~al.(2021)Callebaut, Leenders, Van~Mulders, Ottoy, De~Strycker and Van~der Perre}]{callebaut_art_2021}
Callebaut, G., Leenders, G., Van~Mulders, J., Ottoy, G., De~Strycker, L. and Van~der Perre, L. (2021) The {Art} of {Designing} {Remote} {IoT} {Devices}—{Technologies} and {Strategies} for a {Long} {Battery} {Life}.
\newblock \textit{Sensors}, \textbf{21}, 913.
\newblock \urlprefix\url{https://www.mdpi.com/1424-8220/21/3/913}.
\newblock Number: 3 Publisher: Multidisciplinary Digital Publishing Institute.

\bibitem[{CBD(2022)}]{cbd_kunming-montreal_2022}
CBD, U. (2022) Kunming-{Montreal} {Global} {Biodiversity} {Framework}.
\newblock \textit{Fifteenth meeting of the {Conference} of the {Parties} to the {Convention} on {Biological} {Diversity} ({Part} {Two}) {Decision} 15/4.}, UN Environment Programme, Montreal, Canada.

\bibitem[{Chen et~al.(2024)Chen, Wang, Lin, Lee, Liu, Wu, Hsu, Yang and Jiang}]{chen_machine_2024}
Chen, S.-H., Wang, J.-C., Lin, H.-J., Lee, M.-H., Liu, A.-C., Wu, Y.-L., Hsu, P.-S., Yang, E.-C. and Jiang, J.-A. (2024) A machine learning-based multiclass classification model for bee colony anomaly identification using an {IoT}-based audio monitoring system with an edge computing framework.
\newblock \textit{Expert Systems with Applications}, \textbf{255}, 124898.
\newblock \urlprefix\url{https://www.sciencedirect.com/science/article/pii/S0957417424017652}.

\bibitem[{Coffey et~al.(2019)Coffey, Marx and Neumaier}]{coffey_deepsqueak_2019}
Coffey, K.~R., Marx, R.~E. and Neumaier, J.~F. (2019) {DeepSqueak}: a deep learning-based system for detection and analysis of ultrasonic vocalizations.
\newblock \textit{Neuropsychopharmacology}, \textbf{44}, 859--868.
\newblock \urlprefix\url{https://www.nature.com/articles/s41386-018-0303-6}.
\newblock Publisher: Nature Publishing Group.

\bibitem[{Cretois et~al.(2022)Cretois, Rosten and Sethi}]{cretois_voice_2022}
Cretois, B., Rosten, C.~M. and Sethi, S.~S. (2022) Voice activity detection in eco-acoustic data enables privacy protection and is a proxy for human disturbance.
\newblock \textit{Methods in Ecology and Evolution}, \textbf{13}, 2865--2874.
\newblock \urlprefix\url{https://onlinelibrary.wiley.com/doi/abs/10.1111/2041-210X.14005}.
\newblock \_eprint: https://onlinelibrary.wiley.com/doi/pdf/10.1111/2041-210X.14005.

\bibitem[{Darras et~al.(2024)Darras, Balle, Xu, Yan, Zakka, Toledo-Hernández, Sheng, Lin, Zhang, Lan, Fupeng and Wanger}]{darras_eyes_2024}
Darras, K. F.~A., Balle, M., Xu, W., Yan, Y., Zakka, V.~G., Toledo-Hernández, M., Sheng, D., Lin, W., Zhang, B., Lan, Z., Fupeng, L. and Wanger, T.~C. (2024) Eyes on nature: {Embedded} vision cameras for terrestrial biodiversity monitoring.
\newblock \textit{Methods in Ecology and Evolution}, \textbf{15}, 2262--2275.
\newblock \urlprefix\url{https://onlinelibrary.wiley.com/doi/abs/10.1111/2041-210X.14436}.
\newblock 2784 citations (Semantic Scholar/DOI) [2025-01-23] \_eprint: https://onlinelibrary.wiley.com/doi/pdf/10.1111/2041-210X.14436.

\bibitem[{Desislavov et~al.(2023)Desislavov, Martínez-Plumed and Hernández-Orallo}]{desislavov_trends_2023}
Desislavov, R., Martínez-Plumed, F. and Hernández-Orallo, J. (2023) Trends in {AI} inference energy consumption: {Beyond} the performance-vs-parameter laws of deep learning.
\newblock \textit{Sustainable Computing: Informatics and Systems}, \textbf{38}, 100857.
\newblock \urlprefix\url{https://www.sciencedirect.com/science/article/pii/S2210537923000124}.
\newblock 0 citations (Semantic Scholar/DOI) [2025-01-23].

\bibitem[{Disabato et~al.(2021)Disabato, Canonaco, Flikkema, Roveri and Alippi}]{disabato_birdsong_2021}
Disabato, S., Canonaco, G., Flikkema, P.~G., Roveri, M. and Alippi, C. (2021) Birdsong {Detection} at the {Edge} with {Deep} {Learning}.
\newblock In \textit{2021 {IEEE} {International} {Conference} on {Smart} {Computing} ({SMARTCOMP})}, 9--16.
\newblock ISSN: 2693-8340.

\bibitem[{Fairbrass et~al.(2019)Fairbrass, Firman, Williams, Brostow, Titheridge and Jones}]{fairbrass_citynet-deep_2019}
Fairbrass, A., Firman, M., Williams, C., Brostow, G., Titheridge, H. and Jones, K. (2019) {CityNet}-{Deep} learning tools for urban ecoacoustic assessment.
\newblock \textit{METHODS IN ECOLOGY AND EVOLUTION}, \textbf{10}, 186--197.

\bibitem[{Gallacher et~al.(2021)Gallacher, Wilson, Fairbrass, Turmukhambetov, Firman, Kreitmayer, Mac~Aodha, Brostow and Jones}]{gallacher_shazam_2021}
Gallacher, S., Wilson, D., Fairbrass, A., Turmukhambetov, D., Firman, M., Kreitmayer, S., Mac~Aodha, O., Brostow, G. and Jones, K.~E. (2021) Shazam for bats: {Internet} of {Things} for continuous real-time biodiversity monitoring.
\newblock \textit{IET Smart Cities}, \textbf{3}, 171--183.
\newblock \urlprefix\url{https://onlinelibrary.wiley.com/doi/abs/10.1049/smc2.12016}.
\newblock \_eprint: https://onlinelibrary.wiley.com/doi/pdf/10.1049/smc2.12016.

\bibitem[{Ghani et~al.(2023)Ghani, Denton, Kahl and Klinck}]{ghani_global_2023}
Ghani, B., Denton, T., Kahl, S. and Klinck, H. (2023) Global birdsong embeddings enable superior transfer learning for bioacoustic classification.
\newblock \textit{Scientific Reports}, \textbf{13}, 22876.
\newblock \urlprefix\url{https://www.nature.com/articles/s41598-023-49989-z}.
\newblock 1 citations (Semantic Scholar/DOI) [2025-01-23] Publisher: Nature Publishing Group.

\bibitem[{Gibb et~al.(2019)Gibb, Browning, Glover-Kapfer and Jones}]{gibb_emerging_2019}
Gibb, R., Browning, E., Glover-Kapfer, P. and Jones, K.~E. (2019) Emerging opportunities and challenges for passive acoustics in ecological assessment and monitoring.
\newblock \textit{Methods in Ecology and Evolution}, \textbf{10}, 169--185.
\newblock \urlprefix\url{https://onlinelibrary.wiley.com/doi/abs/10.1111/2041-210X.13101}.
\newblock \_eprint: https://onlinelibrary.wiley.com/doi/pdf/10.1111/2041-210X.13101.

\bibitem[{Gou et~al.(2021)Gou, Yu, Maybank and Tao}]{gou_knowledge_2021}
Gou, J., Yu, B., Maybank, S.~J. and Tao, D. (2021) Knowledge {Distillation}: {A} {Survey}.
\newblock \textit{International Journal of Computer Vision}, \textbf{129}, 1789--1819.
\newblock \urlprefix\url{https://doi.org/10.1007/s11263-021-01453-z}.

\bibitem[{Heath et~al.(2024)Heath, Suzuki, Le~Penru, Skinner, Orme, Ewers, Sethi and Picinali}]{heath_spatial_2024}
Heath, B.~E., Suzuki, R., Le~Penru, N.~P., Skinner, J., Orme, C. D.~L., Ewers, R.~M., Sethi, S.~S. and Picinali, L. (2024) Spatial ecosystem monitoring with a {Multichannel} {Acoustic} {Autonomous} {Recording} {Unit} ({MAARU}).
\newblock \textit{Methods in Ecology and Evolution}, \textbf{15}, 1568--1579.
\newblock \urlprefix\url{https://onlinelibrary.wiley.com/doi/abs/10.1111/2041-210X.14390}.
\newblock \_eprint: https://onlinelibrary.wiley.com/doi/pdf/10.1111/2041-210X.14390.

\bibitem[{Hill et~al.(2019)Hill, Prince, Snaddon, Doncaster and Rogers}]{hill_audiomoth_2019}
Hill, A.~P., Prince, P., Snaddon, J.~L., Doncaster, C.~P. and Rogers, A. (2019) {AudioMoth}: {A} low-cost acoustic device for monitoring biodiversity and the environment.
\newblock \textit{HardwareX}, \textbf{6}, e00073.
\newblock \urlprefix\url{https://www.sciencedirect.com/science/article/pii/S2468067219300306}.

\bibitem[{Hoeﬂer et~al.(2021)Hoeﬂer, Alistarh, Ben-Nun and Dryden}]{hoeer_sparsity_2021}
Hoeﬂer, T., Alistarh, D., Ben-Nun, T. and Dryden, N. (2021) Sparsity in {Deep} {Learning}: {Pruning} and growth for eﬃcient inference and training in neural networks.
\newblock \textit{Journal of Machine Learning Research}, \textbf{22}.
\newblock \urlprefix\url{http://jmlr.org/papers/v22/21-0366.html}.

\bibitem[{Horton et~al.(2019)Horton, Nilsson, Van~Doren, La~Sorte, Dokter and Farnsworth}]{horton_bright_2019}
Horton, K.~G., Nilsson, C., Van~Doren, B.~M., La~Sorte, F.~A., Dokter, A.~M. and Farnsworth, A. (2019) Bright lights in the big cities: migratory birds’ exposure to artificial light.
\newblock \textit{Frontiers in Ecology and the Environment}, \textbf{17}, 209--214.
\newblock \urlprefix\url{https://onlinelibrary.wiley.com/doi/abs/10.1002/fee.2029}.
\newblock \_eprint: https://onlinelibrary.wiley.com/doi/pdf/10.1002/fee.2029.

\bibitem[{Hua et~al.(2023)Hua, Li, Wang, Dong, Li and Cao}]{hua_edge_2023}
Hua, H., Li, Y., Wang, T., Dong, N., Li, W. and Cao, J. (2023) Edge {Computing} with {Artificial} {Intelligence}: {A} {Machine} {Learning} {Perspective}.
\newblock \textit{ACM Comput. Surv.}, \textbf{55}, 184:1--184:35.
\newblock \urlprefix\url{https://dl.acm.org/doi/10.1145/3555802}.
\newblock 45 citations (Semantic Scholar/DOI) [2025-01-23].

\bibitem[{Höchst et~al.(2022)Höchst, Bellafkir, Lampe, Vogelbacher, Mühling, Schneider, Lindner, Rösner, Schabo, Farwig and Freisleben}]{hochst_birdedge_2022}
Höchst, J., Bellafkir, H., Lampe, P., Vogelbacher, M., Mühling, M., Schneider, D., Lindner, K., Rösner, S., Schabo, D.~G., Farwig, N. and Freisleben, B. (2022) Bird@{Edge}: {Bird} {Species} {Recognition} at the {Edge}.
\newblock In \textit{Networked {Systems}} (eds. M.-A. Koulali and M.~Mezini), 69--86. Cham: Springer International Publishing.

\bibitem[{IPBES(2019)}]{ipbes_global_2019}
IPBES (2019) Global assessment report on biodiversity and ecosystem services of the {Intergovernmental} {Science}-{Policy} {Platform} on {Biodiversity} and {Ecosystem} {Services}.
\newblock \textit{Tech. rep.}, Zenodo, Bonn, Germany.
\newblock \urlprefix\url{https://zenodo.org/doi/10.5281/zenodo.3831673}.
\newblock Version Number: 1.

\bibitem[{Jolles(2021)}]{jolles_broad-scale_2021}
Jolles, J.~W. (2021) Broad-scale applications of the {Raspberry} {Pi}: {A} review and guide for biologists.
\newblock \textit{Methods in Ecology and Evolution}, \textbf{12}, 1562--1579.
\newblock \urlprefix\url{https://onlinelibrary.wiley.com/doi/abs/10.1111/2041-210X.13652}.
\newblock \_eprint: https://onlinelibrary.wiley.com/doi/pdf/10.1111/2041-210X.13652.

\bibitem[{Jones and Holderied(2007)}]{jones_bat_2007}
Jones, G. and Holderied, M.~W. (2007) Bat echolocation calls: adaptation and convergent evolution.
\newblock \textit{Proceedings of the Royal Society B: Biological Sciences}.
\newblock \urlprefix\url{https://royalsocietypublishing.org/doi/10.1098/rspb.2006.0200}.
\newblock Publisher: The Royal SocietyLondon.

\bibitem[{Kahl et~al.(2021)Kahl, Wood, Eibl and Klinck}]{kahl_birdnet_2021}
Kahl, S., Wood, C.~M., Eibl, M. and Klinck, H. (2021) {BirdNET}: {A} deep learning solution for avian diversity monitoring.
\newblock \textit{Ecological Informatics}, \textbf{61}, 101236.
\newblock \urlprefix\url{https://www.sciencedirect.com/science/article/pii/S1574954121000273}.

\bibitem[{Karlsson et~al.(2021)Karlsson, Tay, Imbun and Hughes}]{karlsson_kinabalu_2021}
Karlsson, E. C.~M., Tay, H., Imbun, P. and Hughes, A.~C. (2021) The {Kinabalu} {Recorder}, a new passive acoustic and environmental monitoring recorder.
\newblock \textit{Methods in Ecology and Evolution}, \textbf{12}, 2109--2116.
\newblock \urlprefix\url{https://onlinelibrary.wiley.com/doi/abs/10.1111/2041-210X.13671}.
\newblock \_eprint: https://onlinelibrary.wiley.com/doi/pdf/10.1111/2041-210X.13671.

\bibitem[{Kim et~al.(2022)Kim, Park and Lu}]{kim_survey_2022}
Kim, S., Park, K.-J. and Lu, C. (2022) A {Survey} on {Network} {Security} for {Cyber}–{Physical} {Systems}: {From} {Threats} to {Resilient} {Design}.
\newblock \textit{IEEE Communications Surveys \& Tutorials}, \textbf{24}, 1534--1573.
\newblock \urlprefix\url{https://ieeexplore.ieee.org/document/9810983/?arnumber=9810983}.
\newblock Conference Name: IEEE Communications Surveys \& Tutorials.

\bibitem[{Knight et~al.(2020)Knight, Sòlymos, Scott and Bayne}]{knight_validation_2020}
Knight, E.~C., Sòlymos, P., Scott, C. and Bayne, E.~M. (2020) Validation prediction: a flexible protocol to increase efficiency of automated acoustic processing for wildlife research.
\newblock \textit{Ecological Applications}, \textbf{30}, e02140.
\newblock \urlprefix\url{https://onlinelibrary.wiley.com/doi/abs/10.1002/eap.2140}.
\newblock \_eprint: https://onlinelibrary.wiley.com/doi/pdf/10.1002/eap.2140.

\bibitem[{Krishnamoorthi(2018)}]{krishnamoorthi_quantizing_2018}
Krishnamoorthi, R. (2018) Quantizing deep convolutional networks for efficient inference: {A} whitepaper.
\newblock \urlprefix\url{http://arxiv.org/abs/1806.08342}.
\newblock ArXiv:1806.08342 [cs].

\bibitem[{Lapp et~al.(2023)Lapp, Rhinehart, Freeland-Haynes, Khilnani, Syunkova and Kitzes}]{lapp_opensoundscape_2023}
Lapp, S., Rhinehart, T., Freeland-Haynes, L., Khilnani, J., Syunkova, A. and Kitzes, J. (2023) {OpenSoundscape}: {An} open-source bioacoustics analysis package for {Python}.
\newblock \textit{Methods in Ecology and Evolution}, \textbf{14}, 2321--2328.
\newblock \urlprefix\url{https://onlinelibrary.wiley.com/doi/abs/10.1111/2041-210X.14196}.
\newblock \_eprint: https://onlinelibrary.wiley.com/doi/pdf/10.1111/2041-210X.14196.

\bibitem[{Lapp et~al.(2021)Lapp, Wu, Richards-Zawacki, Voyles, Rodriguez, Shamon and Kitzes}]{lapp_automated_2021}
Lapp, S., Wu, T., Richards-Zawacki, C., Voyles, J., Rodriguez, K.~M., Shamon, H. and Kitzes, J. (2021) Automated detection of frog calls and choruses by pulse repetition rate.
\newblock \textit{Conservation Biology}, \textbf{35}, 1659--1668.
\newblock \urlprefix\url{https://onlinelibrary.wiley.com/doi/abs/10.1111/cobi.13718}.
\newblock \_eprint: https://onlinelibrary.wiley.com/doi/pdf/10.1111/cobi.13718.

\bibitem[{Mac~Aodha et~al.(2022)Mac~Aodha, Martínez~Balvanera, Damstra, Cooke, Eichinski, Browning, Barataud, Boughey, Coles, Giacomini, Mac Swiney~G., Obrist, Parsons, Sattler and Jones}]{mac_aodha_towards_2022}
Mac~Aodha, O., Martínez~Balvanera, S., Damstra, E., Cooke, M., Eichinski, P., Browning, E., Barataud, M., Boughey, K., Coles, R., Giacomini, G., Mac Swiney~G., M.~C., Obrist, M.~K., Parsons, S., Sattler, T. and Jones, K.~E. (2022) Towards a {General} {Approach} for {Bat} {Echolocation} {Detection} and {Classification}.
\newblock \urlprefix\url{https://www.biorxiv.org/content/10.1101/2022.12.14.520490v1}.
\newblock Pages: 2022.12.14.520490 Section: New Results.

\bibitem[{Martínez~Balvanera et~al.(2024)Martínez~Balvanera, Mac~Aodha, Weldy, Pringle, Browning and Jones}]{martinez_balvanera_whombat_2024}
Martínez~Balvanera, S., Mac~Aodha, O., Weldy, M.~J., Pringle, H., Browning, E. and Jones, K.~E. (2024) Whombat: {An} open-source audio annotation tool for machine learning assisted bioacoustics.
\newblock \textit{Methods in Ecology and Evolution}, \textbf{n/a}.
\newblock \urlprefix\url{https://onlinelibrary.wiley.com/doi/abs/10.1111/2041-210X.14468}.
\newblock \_eprint: https://onlinelibrary.wiley.com/doi/pdf/10.1111/2041-210X.14468.

\bibitem[{McGuire(2023)}]{mcguire_birdnet-pi_2023}
McGuire, P. (2023) {BirdNET}-{Pi}.
\newblock \urlprefix\url{https://github.com/mcguirepr89/BirdNET-Pi}.
\newblock Original-date: 2021-09-28T17:41:36Z.

\bibitem[{van Merriënboer et~al.(2024)van Merriënboer, Hamer, Dumoulin, Triantafillou and Denton}]{van_merrienboer_birds_2024}
van Merriënboer, B., Hamer, J., Dumoulin, V., Triantafillou, E. and Denton, T. (2024) Birds, bats and beyond: evaluating generalization in bioacoustics models.
\newblock \textit{Frontiers in Bird Science}, \textbf{3}.
\newblock \urlprefix\url{https://www.frontiersin.org/journals/bird-science/articles/10.3389/fbirs.2024.1369756/full}.
\newblock Publisher: Frontiers.

\bibitem[{Metcalf et~al.(2023)Metcalf, Abrahams, Ashington, Baker, Bradfer-Lawrence, Browning, Carruthers-Jones, Darby, Dick, Eldridge, Elliot, Heath, Howden-Leach, Johnston, Lees, Meyer, Ruiz~Arana and Smyth}]{metcalf_good_2023}
Metcalf, O., Abrahams, C., Ashington, B., Baker, B., Bradfer-Lawrence, T., Browning, E., Carruthers-Jones, J., Darby, J., Dick, J., Eldridge, A., Elliot, D., Heath, B., Howden-Leach, P., Johnston, A., Lees, A., Meyer, C., Ruiz~Arana, U. and Smyth, S. (2023) Good practice guidelines for long-term ecoacoustic monitoring in the {UK}.
\newblock \textit{Educational material}, UK Acoustics Network.

\bibitem[{Millar et~al.(2024)Millar, Sethi, Haddadi and Madhavapeddy}]{millar_terracorder_2024}
Millar, J., Sethi, S., Haddadi, H. and Madhavapeddy, A. (2024) Terracorder: {Sense} {Long} and {Prosper}.
\newblock \urlprefix\url{http://arxiv.org/abs/2408.02407}.
\newblock ArXiv:2408.02407 [cs].

\bibitem[{Mooney et~al.(2020)Mooney, Di~Iorio, Lammers, Lin, Nedelec, Parsons, Radford, Urban and Stanley}]{mooney_listening_2020}
Mooney, T.~A., Di~Iorio, L., Lammers, M., Lin, T.-H., Nedelec, S.~L., Parsons, M., Radford, C., Urban, E. and Stanley, J. (2020) Listening forward: approaching marine biodiversity assessments using acoustic methods.
\newblock \textit{Royal Society Open Science}, \textbf{7}, 201287.
\newblock \urlprefix\url{https://royalsocietypublishing.org/doi/10.1098/rsos.201287}.
\newblock Publisher: Royal Society.

\bibitem[{Napier et~al.(2024)Napier, Ahn, Allen-Ankins, Schwarzkopf and Lee}]{napier_advancements_2024}
Napier, T., Ahn, E., Allen-Ankins, S., Schwarzkopf, L. and Lee, I. (2024) Advancements in preprocessing, detection and classification techniques for ecoacoustic data: {A} comprehensive review for large-scale {Passive} {Acoustic} {Monitoring}.
\newblock \textit{Expert Systems with Applications}, \textbf{252}, 124220.
\newblock \urlprefix\url{https://linkinghub.elsevier.com/retrieve/pii/S0957417424010868}.

\bibitem[{Navine et~al.(2024)Navine, Denton, Weldy and Hart}]{navine_all_2024}
Navine, A.~K., Denton, T., Weldy, M.~J. and Hart, P.~J. (2024) All thresholds barred: direct estimation of call density in bioacoustic data.
\newblock \textit{Frontiers in Bird Science}, \textbf{3}, 1380636.
\newblock \urlprefix\url{https://www.frontiersin.org/articles/10.3389/fbirs.2024.1380636/full}.
\newblock 0 citations (Semantic Scholar/DOI) [2025-01-23].

\bibitem[{Novac et~al.(2021)Novac, Boukli~Hacene, Pegatoquet, Miramond and Gripon}]{novac_quantization_2021}
Novac, P.-E., Boukli~Hacene, G., Pegatoquet, A., Miramond, B. and Gripon, V. (2021) Quantization and {Deployment} of {Deep} {Neural} {Networks} on {Microcontrollers}.
\newblock \textit{Sensors}, \textbf{21}, 2984.
\newblock \urlprefix\url{https://www.mdpi.com/1424-8220/21/9/2984}.
\newblock Number: 9 Publisher: Multidisciplinary Digital Publishing Institute.

\bibitem[{Pérez-Granados(2023)}]{perez-granados_birdnet_2023}
Pérez-Granados, C. (2023) {BirdNET}: applications, performance, pitfalls and future opportunities.
\newblock \textit{Ibis}, \textbf{165}, 1068--1075.
\newblock \urlprefix\url{https://onlinelibrary.wiley.com/doi/abs/10.1111/ibi.13193}.
\newblock 1158 citations (Semantic Scholar/DOI) [2025-01-23] \_eprint: https://onlinelibrary.wiley.com/doi/pdf/10.1111/ibi.13193.

\bibitem[{Rhinehart et~al.(2020)Rhinehart, Chronister, Devlin and Kitzes}]{rhinehart_acoustic_2020}
Rhinehart, T.~A., Chronister, L.~M., Devlin, T. and Kitzes, J. (2020) Acoustic localization of terrestrial wildlife: {Current} practices and future opportunities.
\newblock \textit{Ecology and Evolution}, \textbf{10}, 6794--6818.
\newblock \urlprefix\url{https://onlinelibrary.wiley.com/doi/abs/10.1002/ece3.6216}.
\newblock \_eprint: https://onlinelibrary.wiley.com/doi/pdf/10.1002/ece3.6216.

\bibitem[{Rhinehart et~al.(2022)Rhinehart, Turek and Kitzes}]{rhinehart_continuous-score_2022}
Rhinehart, T.~A., Turek, D. and Kitzes, J. (2022) A continuous-score occupancy model that incorporates uncertain machine learning output from autonomous biodiversity surveys.
\newblock \textit{Methods in Ecology and Evolution}, \textbf{13}, 1778--1789.
\newblock \urlprefix\url{https://onlinelibrary.wiley.com/doi/abs/10.1111/2041-210X.13905}.
\newblock \_eprint: https://onlinelibrary.wiley.com/doi/pdf/10.1111/2041-210X.13905.

\bibitem[{Roe et~al.(2021)Roe, Eichinski, Fuller, McDonald, Schwarzkopf, Towsey, Truskinger, Tucker and Watson}]{roe_australian_2021}
Roe, P., Eichinski, P., Fuller, R.~A., McDonald, P.~G., Schwarzkopf, L., Towsey, M., Truskinger, A., Tucker, D. and Watson, D.~M. (2021) The {Australian} {Acoustic} {Observatory}.
\newblock \textit{Methods in Ecology and Evolution}, \textbf{12}, 1802--1808.
\newblock \urlprefix\url{https://onlinelibrary.wiley.com/doi/abs/10.1111/2041-210X.13660}.
\newblock \_eprint: https://onlinelibrary.wiley.com/doi/pdf/10.1111/2041-210X.13660.

\bibitem[{Ross et~al.(2021)Ross, Friedman, Yoshimura, Yoshida, Donohue and Economo}]{ross_utility_2021}
Ross, S. R. P.-J., Friedman, N.~R., Yoshimura, M., Yoshida, T., Donohue, I. and Economo, E.~P. (2021) Utility of acoustic indices for ecological monitoring in complex sonic environments.
\newblock \textit{Ecological Indicators}, \textbf{121}, 107114.
\newblock \urlprefix\url{https://www.sciencedirect.com/science/article/pii/S1470160X20310530}.

\bibitem[{Ross et~al.(2023)Ross, O'Connell, Deichmann, Desjonquères, Gasc, Phillips, Sethi, Wood and Burivalova}]{ross_passive_2023}
Ross, S. R. P.-J., O'Connell, D.~P., Deichmann, J.~L., Desjonquères, C., Gasc, A., Phillips, J.~N., Sethi, S.~S., Wood, C.~M. and Burivalova, Z. (2023) Passive acoustic monitoring provides a fresh perspective on fundamental ecological questions.
\newblock \textit{Functional Ecology}, \textbf{37}, 959--975.
\newblock \urlprefix\url{https://onlinelibrary.wiley.com/doi/abs/10.1111/1365-2435.14275}.
\newblock \_eprint: https://onlinelibrary.wiley.com/doi/pdf/10.1111/1365-2435.14275.

\bibitem[{Sandbrook et~al.(2021)Sandbrook, Clark, Toivonen, Simlai, O'Donnell, Cobbe and Adams}]{sandbrook_principles_2021}
Sandbrook, C., Clark, D., Toivonen, T., Simlai, T., O'Donnell, S., Cobbe, J. and Adams, W. (2021) Principles for the socially responsible use of conservation monitoring technology and data.
\newblock \textit{Conservation Science and Practice}, \textbf{3}, e374.
\newblock \urlprefix\url{https://onlinelibrary.wiley.com/doi/abs/10.1111/csp2.374}.
\newblock \_eprint: https://onlinelibrary.wiley.com/doi/pdf/10.1111/csp2.374.

\bibitem[{Sethi et~al.(2024)Sethi, Bick, Chen, Crouzeilles, Hillier, Lawson, Lee, Liu, de~Freitas~Parruco, Rosten, Somveille, Tuanmu and Banks-Leite}]{sethi_large-scale_2024}
Sethi, S.~S., Bick, A., Chen, M.-Y., Crouzeilles, R., Hillier, B.~V., Lawson, J., Lee, C.-Y., Liu, S.-H., de~Freitas~Parruco, C.~H., Rosten, C.~M., Somveille, M., Tuanmu, M.-N. and Banks-Leite, C. (2024) Large-scale avian vocalization detection delivers reliable global biodiversity insights.
\newblock \textit{Proceedings of the National Academy of Sciences}, \textbf{121}, e2315933121.
\newblock \urlprefix\url{https://www.pnas.org/doi/10.1073/pnas.2315933121}.
\newblock 2 citations (Semantic Scholar/DOI) [2024-11-29] Publisher: Proceedings of the National Academy of Sciences.

\bibitem[{Sethi et~al.(2018)Sethi, Ewers, Jones, Orme and Picinali}]{sethi_robust_2018}
Sethi, S.~S., Ewers, R.~M., Jones, N.~S., Orme, C. D.~L. and Picinali, L. (2018) Robust, real-time and autonomous monitoring of ecosystems with an open, low-cost, networked device.
\newblock \textit{Methods in Ecology and Evolution}, \textbf{9}, 2383--2387.
\newblock \urlprefix\url{https://onlinelibrary.wiley.com/doi/abs/10.1111/2041-210X.13089}.
\newblock \_eprint: https://onlinelibrary.wiley.com/doi/pdf/10.1111/2041-210X.13089.

\bibitem[{Sethi et~al.(2020)Sethi, Ewers, Jones, Signorelli, Picinali and Orme}]{sethi_safe_2020}
Sethi, S.~S., Ewers, R.~M., Jones, N.~S., Signorelli, A., Picinali, L. and Orme, C. D.~L. (2020) {SAFE} {Acoustics}: {An} open-source, real-time eco-acoustic monitoring network in the tropical rainforests of {Borneo}.
\newblock \textit{Methods in Ecology and Evolution}, \textbf{11}, 1182--1185.
\newblock \urlprefix\url{https://onlinelibrary.wiley.com/doi/abs/10.1111/2041-210X.13438}.
\newblock \_eprint: https://onlinelibrary.wiley.com/doi/pdf/10.1111/2041-210X.13438.

\bibitem[{Solem et~al.(2025)Solem, Uddin and Nosrati}]{solem_celery_2025}
Solem, A.~H., Uddin, A.~S. and Nosrati, T. (2025) Celery.
\newblock \urlprefix\url{https://github.com/celery/celery}.
\newblock Original-date: 2009-04-24T11:31:24Z.

\bibitem[{Stephenson et~al.(2022)Stephenson, Londoño-Murcia, Borges, Claassens, Frisch-Nwakanma, Ling, McMullan-Fisher, Meeuwig, Unter, Walls, Burfield, do~Carmo Vieira~Correa, Geller, Montenegro~Paredes, Mubalama, Ntiamoa-Baidu, Roesler, Rovero, Sharma, Wiwardhana, Yang and Fumagalli}]{stephenson_measuring_2022}
Stephenson, P.~J., Londoño-Murcia, M.~C., Borges, P. A.~V., Claassens, L., Frisch-Nwakanma, H., Ling, N., McMullan-Fisher, S., Meeuwig, J.~J., Unter, K. M.~M., Walls, J.~L., Burfield, I.~J., do~Carmo Vieira~Correa, D., Geller, G.~N., Montenegro~Paredes, I., Mubalama, L.~K., Ntiamoa-Baidu, Y., Roesler, I., Rovero, F., Sharma, Y.~P., Wiwardhana, N.~W., Yang, J. and Fumagalli, L. (2022) Measuring the {Impact} of {Conservation}: {The} {Growing} {Importance} of {Monitoring} {Fauna}, {Flora} and {Funga}.
\newblock \textit{Diversity}, \textbf{14}, 824.
\newblock \urlprefix\url{https://www.mdpi.com/1424-2818/14/10/824}.
\newblock Number: 10 Publisher: Multidisciplinary Digital Publishing Institute.

\bibitem[{Stowell(2022)}]{stowell_computational_2022}
Stowell, D. (2022) Computational bioacoustics with deep learning: a review and roadmap.
\newblock \textit{PeerJ}, \textbf{10}, e13152.
\newblock \urlprefix\url{https://peerj.com/articles/13152}.
\newblock Publisher: PeerJ Inc.

\bibitem[{Stähli et~al.(2022)Stähli, Ost and Studer}]{stahli_development_2022}
Stähli, O., Ost, T. and Studer, T. (2022) Development of an {AI}-based bioacoustic wolf monitoring system.
\newblock \textit{The International FLAIRS Conference Proceedings}, \textbf{35}.
\newblock \urlprefix\url{https://journals.flvc.org/FLAIRS/article/view/130552}.

\bibitem[{Sugai et~al.(2019)Sugai, Silva, Ribeiro and Llusia}]{sugai_terrestrial_2019}
Sugai, L. S.~M., Silva, T. S.~F., Ribeiro, Jr, J.~W. and Llusia, D. (2019) Terrestrial {Passive} {Acoustic} {Monitoring}: {Review} and {Perspectives}.
\newblock \textit{BioScience}, \textbf{69}, 15--25.
\newblock \urlprefix\url{https://doi.org/10.1093/biosci/biy147}.

\bibitem[{Teixeira et~al.(2024)Teixeira, Roe, van Rensburg, Linke, McDonald, Tucker and Fuller}]{teixeira_effective_2024}
Teixeira, D., Roe, P., van Rensburg, B.~J., Linke, S., McDonald, P.~G., Tucker, D. and Fuller, S. (2024) Effective ecological monitoring using passive acoustic sensors: {Recommendations} for conservation practitioners.
\newblock \textit{Conservation Science and Practice}, \textbf{6}, e13132.
\newblock \urlprefix\url{https://onlinelibrary.wiley.com/doi/abs/10.1111/csp2.13132}.
\newblock \_eprint: https://onlinelibrary.wiley.com/doi/pdf/10.1111/csp2.13132.

\bibitem[{Ulloa et~al.(2021)Ulloa, Haupert, Latorre, Aubin and Sueur}]{ulloa_scikit-maad_2021}
Ulloa, J.~S., Haupert, S., Latorre, J.~F., Aubin, T. and Sueur, J. (2021) scikit-maad: {An} open-source and modular toolbox for quantitative soundscape analysis in {Python}.
\newblock \textit{Methods in Ecology and Evolution}, \textbf{12}, 2334--2340.
\newblock \urlprefix\url{https://onlinelibrary.wiley.com/doi/abs/10.1111/2041-210X.13711}.
\newblock \_eprint: https://onlinelibrary.wiley.com/doi/pdf/10.1111/2041-210X.13711.

\bibitem[{Whytock and Christie(2017)}]{whytock_solo_2017}
Whytock, R.~C. and Christie, J. (2017) Solo: an open source, customizable and inexpensive audio recorder for bioacoustic research.
\newblock \textit{Methods in Ecology and Evolution}, \textbf{8}, 308--312.
\newblock \urlprefix\url{https://onlinelibrary.wiley.com/doi/abs/10.1111/2041-210X.12678}.
\newblock \_eprint: https://onlinelibrary.wiley.com/doi/pdf/10.1111/2041-210X.12678.

\bibitem[{Wood et~al.(2024)Wood, Günther, Rex, Hofstadter, Reers, Kahl, Peery and Klinck}]{wood_real-time_2024}
Wood, C.~M., Günther, F., Rex, A., Hofstadter, D.~F., Reers, H., Kahl, S., Peery, M.~Z. and Klinck, H. (2024) Real-time acoustic monitoring facilitates the proactive management of biological invasions.
\newblock \textit{Biological Invasions}.
\newblock \urlprefix\url{https://link.springer.com/10.1007/s10530-024-03426-y}.

\bibitem[{Zook et~al.(2017)Zook, Barocas, Boyd, Crawford, Keller, Gangadharan, Goodman, Hollander, Koenig, Metcalf, Narayanan, Nelson and Pasquale}]{zook_ten_2017}
Zook, M., Barocas, S., Boyd, D., Crawford, K., Keller, E., Gangadharan, S.~P., Goodman, A., Hollander, R., Koenig, B.~A., Metcalf, J., Narayanan, A., Nelson, A. and Pasquale, F. (2017) Ten simple rules for responsible big data research.
\newblock \textit{PLOS Computational Biology}, \textbf{13}, e1005399.
\newblock \urlprefix\url{https://journals.plos.org/ploscompbiol/article?id=10.1371/journal.pcbi.1005399}.
\newblock Publisher: Public Library of Science.

\bibitem[{Zualkernan et~al.(2021)Zualkernan, Judas, Mahbub, Bhagwagar and Chand}]{zualkernan_aiot_2021}
Zualkernan, I., Judas, J., Mahbub, T., Bhagwagar, A. and Chand, P. (2021) An {AIoT} {System} for {Bat} {Species} {Classification}.
\newblock In \textit{2020 {IEEE} {International} {Conference} on {Internet} of {Things} and {Intelligence} {System} ({IoTaIS})}, 155--160.

\end{thebibliography}

\clearpage
\appendix{}
\section{Appendix}
\subsection{Examples Bioacoustic Research Projects}
\label{appendix:bioacoustic_table}

\begin{table}[H]
    \centering
    \caption{Quantity and size of audio recording for various acoustic research projects.}
    \renewcommand{\arraystretch}{2} % Adjust row height for better spacing
    \adjustbox{width=\textwidth}{%
    \fontsize{18}{22}\selectfont
    \begin{tabular}{|>
        {\raggedright\arraybackslash}p{5.2cm}|>
        {\raggedright\arraybackslash}p{4.5cm}|>
        {\raggedright\arraybackslash}p{4.5cm}|>
        {\raggedright\arraybackslash}p{4.5cm}|>
        {\raggedright\arraybackslash}p{4.8cm}|>
        {\raggedright\arraybackslash}p{4.8cm}|>
        {\raggedright\arraybackslash}p{4.8cm}|}
        \hline
        \textbf{Project Description} & \textbf{Number of Recording Devices} & \textbf{Sample Rate (Hz)} & \textbf{Total Number of Recordings} & \textbf{Length of 1x Recording (minutes)} & \textbf{Total Recording Duration (hours)} & \textbf{Total Data Size (TB)} \\ \hline
        BiomeHealth, Maasai Mara, Kenya, 2019 & 74 & 384,000 & 87,543 & 1 & 1,459.05 & 32.27 \\ \hline
        Williams B. CoralReef, Pabbiring Islands Indonesia, 2022 & 18 & 16,000 & 267,688 & 1 & 4,461.46 & 0.514 \\ 
        \hline
    \end{tabular}%
    }
    \label{tab:recording_data}
\end{table}

\begin{table}[H]
    \centering
    \caption{Transfer, speed and duration of audio data upload for various acoustic research projects.}
    \renewcommand{\arraystretch}{2.5} % Adjust row height for better spacing
    \adjustbox{width=\textwidth}{%
    \fontsize{20}{24}\selectfont
    \begin{tabular}{|>
        {\raggedright\arraybackslash}p{5.2cm}|>
        {\raggedright\arraybackslash}p{4.5cm}|>
        {\raggedright\arraybackslash}p{5cm}|>
        {\raggedright\arraybackslash}p{4.8cm}|>
        {\raggedright\arraybackslash}p{4.8cm}|>
        {\raggedright\arraybackslash}p{4.8cm}|>
        {\raggedright\arraybackslash}p{4.8cm}|>
        {\raggedright\arraybackslash}p{4.5cm}|}
        \hline
        \textbf{Project Description} & \textbf{Total Data Size (TB)} & \textbf{Transfer Method} & \textbf{SDCard Read Speed} & \textbf{SDCard Data Transfer Time (hours)} & \textbf{Data Storage Location (End Destination)} & \textbf{Data Upload to Server, WiFi Speed (Mbps)} & \textbf{Total Upload Time} \\ \hline
        BiomeHealth, Maasai Mara, Kenya, 2019 & 32.27 & Physical SDCard Transfer Kenya to London & 200MB/s (SanDisk Extreme Plus microSD) & 44.82 & UCL Data Server & 350 & 8d 12h 53min \\ \hline
        Williams B. CoralReef, Pabbiring Islands Indonesia, 2022 & 0.514 & SDCards to Hard Drive. Physical Hard Drive Transfer Indonesia to London & 170MB/s (SanDisk Extreme 64GB microSD) & 6.716 & UCL Data Server, GCP (Google Cloud) & 50 &  22h 50min \\ 
        \hline
    \end{tabular}%
    }
    \label{tab:data_transfer_details}
\end{table}

\subsection{Configuration Parameters of \texttt{acoupi} Deployment}
\label{appendix:configuration_table}

\begin{longtable}{|l|l|l|}
        \caption{Configuration parameters for acoupi\_birdnet and acoupi\_batdetect2 programmes.\label{long}} \\
        \hline
        \textbf{Parameter} & \textbf{Values for acoupi\_birdnet} & \textbf{Values for acoupi\_batdetect2} \\ \hline
    \endfirsthead
        \multicolumn{3}{c}{\tablename\ \thetable\ -- \textit{Continued from previous page}} \\
        \hline
        \textbf{Parameter} & \textbf{Values for acoupi\_birdnet} & \textbf{Values for acoupi\_batdetect2} \\ \hline
    \endhead
        \hline \multicolumn{3}{r}{\textit{Continued on next page}} \\
    \endfoot
        \hline
    \endlastfoot

    \textbf{Timezone} & Europe/London & Europe/London \\ \hline
    \textbf{Microphone} & & \\ \hline
    \hspace{0.2cm}{Microphone Device Name} & UAC 1.0 Microphone \& HID-Mediak & UltraMic 250K 16 bit r4 \\ \hline
    \hspace{0.2cm}{Microphone Samplerate} & 44100 Hz & 250000 Hz \\ \hline
    \hspace{0.2cm}{Microphone Audio Channels} & 1 & 1 \\ \hline
    \textbf{Recording} & & \\ \hline
    \hspace{0.2cm}{Duration} & 9 seconds & 3 seconds \\ \hline
    \hspace{0.2cm}{Interval} & 10 seconds & 10 seconds \\ \hline
    \hspace{0.2cm}{Chunksize} & 8192 & 8192 \\ \hline
    \hspace{0.2cm}{Schedule Start} & 00:00:00 & 17:00:00 \\ \hline
    \hspace{0.2cm}{Schedule End} & 00:00:00 & 07:00:00 \\ \hline
    \textbf{Paths} & & \\ \hline
    \hspace{0.2cm}{Temporary Audio Path} & /run/shm & /run/shm \\ \hline
    \hspace{0.2cm}{Recordings Path} & /home/pi/storages/recordings  & /home/pi/storages/recordings \\ \hline
    \hspace{0.2cm}{Metadata Database Path} & /home/pi/storages/metadata.db & /home/pi/storages/metadata.db \\ \hline
    \textbf{Messaging} & & \\ \hline
    \hspace{0.2cm}{Messages Database Path} & /home/pi/storages/messages.db & /home/pi/storages/messages.db \\ \hline
    \hspace{0.2cm}{Message Send Interval} & 60 seconds & 60 seconds \\ \hline
    \hspace{0.2cm}{Heartbeat Interval} & 3600 seconds  & 3600 seconds \\ \hline
    \hspace{0.2cm}{HTTP} & null & null \\ \hline
    \hspace{0.2cm}{\textbf{MQTT}} & & \\ \hline
    \hspace{0.4cm}{MQTT Host} & \textit{mqtt.yourhost.org} & \textit{mqtt.yourhost.org} \\ \hline
    \hspace{0.4cm}{MQTT Username} & \textit{your-username} & \textit{your-username} \\ \hline
    \hspace{0.4cm}{MQTT Password} & \textit{your-password} & \textit{your-password} \\ \hline
    \hspace{0.4cm}{MQTT Topic} & /acoupi/birds & /acoupi/bats \\ \hline
    \hspace{0.4cm}{MQTT Port} & 1884 & 1884 \\ \hline
    \hspace{0.4cm}{MQTT Timeout} & 5 seconds & 5 seconds \\ \hline
    \textbf{Model} & & \\ \hline
    \textbf{Detection Threshold} & 0.4 & 0.4 \\ \hline
    \textbf{Saving Filters} & & \\ \hline
    \hspace{0.2cm}{Filter Starttime} & 06:00:00 & 17:00:00 \\ \hline
    \hspace{0.2cm}{Filter Endtime} & 22:00:00 & 07:00:00 \\ \hline
    \hspace{0.2cm}{Before Dawn/Dusk Duration} & 30 minutes & 0 \\ \hline
    \hspace{0.2cm}{After Dawn/Dusk Duration} & 30 minutes & 0 \\ \hline
    \hspace{0.2cm}{Frequency Duration} & 0 & 0 \\ \hline
    \hspace{0.2cm}{Frequency Interval} & 0 & 0 \\ \hline
    \textbf{Saving Managers} & & \\ \hline
    \hspace{0.2cm}{True Directory} & birds & bats \\ \hline
    \hspace{0.2cm}{False Directory} & no\_birds & no\_bats \\ \hline
    \hspace{0.2cm}{Saving Time Format} & \%Y\%m\%d\_\%H\%M\%S & \%Y\%m\%d\_\%H\%M\%S \\ \hline
    \hspace{0.2cm}{Saving Threshold} & 0.4 & 0.4 \\ \hline
    \textbf{Summariser Config} & & \\ \hline
    \hspace{0.2cm}{Summary Interval} & 3600.0 seconds & 3600.0 seconds \\ \hline
    \hspace{0.2cm}{Low Band Threshold} & 0.0 & 0.0 \\ \hline
    \hspace{0.2cm}{Mid Band Threshold} & 0.0 & 0.0 \\ \hline
    \hspace{0.2cm}{High Band Threshold} & 0.0 & 0.0 \\ \hline
\end{longtable}

\subsection{Results Summary of \texttt{acoupi} Deployment}
\label{appendix:deployment_results}

\begin{table}[H]
    \centering
    \caption{Summary of Deployment for \textit{acoupi\_birdnet} and \textit{acoupi\_batdetect2}.}
    \renewcommand{\arraystretch}{2} % Adjust row height for better spacing
    \adjustbox{width=\textwidth}{%
    \fontsize{19}{24}\selectfont
    \begin{tabular}
        {|>{\raggedright\arraybackslash}p{4.5cm}|>{\raggedright\arraybackslash}p{4.5cm}|>{\raggedright\arraybackslash}p{4.5cm}|>{\raggedright\arraybackslash}p{4.5cm}|>{\raggedright\arraybackslash}p{4.8cm}|>{\raggedright\arraybackslash}p{4.8cm}|>{\raggedright\arraybackslash}p{4.8cm}|} \hline
        \textbf{programme Name} & \textbf{Expected Nb. Recordings} & \textbf{Actual Nb. Recordings (Coverage \% )} & \textbf{Nb. Recordings Processed} & \textbf{Nb. Recordings Unprocessed} & \textbf{Nb. Messages Sent Successfully} & \textbf{Nb. Messages Sent Failed} \\ \hline
        acoupi\_ birdnet & 130,098 & 129,939 (99.88\%) & 129,935 & 4 & 8716 & 0 \\ \hline
        acoupi\_ batdetect2 & 66,233 & 65,711 (99.21\%) & 64,508 & 1,203 (1.83\%) & 868 & 0\\ \hline
    \end{tabular}%
    }
    \label{tab:acoupi_test_deployment_summary}
\end{table}

\begin{table}[H]
\centering
    \caption{Summary of Detections for\textit{ acoupi\_birdnet} and \textit{acoupi\_batdetect2}.}
    \renewcommand{\arraystretch}{2} % Adjust row height for better spacing
    \adjustbox{width=\textwidth}{%
    \fontsize{19}{24}\selectfont
    \begin{tabular}{|>
        {\raggedright\arraybackslash}p{4.5cm}|>
        {\raggedright\arraybackslash}p{4.3cm}|>
        {\raggedright\arraybackslash}p{4.3cm}|>
        {\raggedright\arraybackslash}p{4.3cm}|>
        {\raggedright\arraybackslash}p{4.8cm}|>
        {\raggedright\arraybackslash}p{4.8cm}|>
        {\raggedright\arraybackslash}p{4.8cm}|}
        \hline
        \textbf{programme Name} & 
        \textbf{Size of DB "metadata"} & 
        \textbf{Size of DB "messages"} & 
        \textbf{Total Nb. Detections} & 
        \textbf{Nb. Detections (Detection Score > 0.4)} & 
        \textbf{Nb. Predicted Tags (Classification Score > 0.4)} & 
        \textbf{Nb. Different Species Class (Classification Score > 0.4)} \\ 
        \hline
        acoupi\_ birdnet & 
        542.3 MB & 
        10.4 MB & 
        23,229 &
        10,838 & 
        10,838 & 
        398 \\ 
        \hline
        acoupi\_ batdetect2 & 
        5.67 GB & 
        1.0 MB & 
        19,416,908 &
        1,031 & 
        174 & 
        4 \\ 
        \hline
    \end{tabular}%
    }
\label{tab:acoupi_test_detections_summariy}
\end{table}

\subsection{Summary of top 20 detection results for \textit{acoupi\_birdnet} and \textit{acoupi\_batdetect2}}
\label{appendix:detection_results}

Although not the primary focus of this study, the detections made by the two bioacoustic classifiers were consistent with the expected soundscape of the deployment location and the seasonality of the test.
The roof garden is urban, close to a busy traffic road, a railway track, but in the proximity of the Waterworks River, where common water birds are found.

Common UK bird species identified with high confidence (score > 0.85) included the Eurasian magpie (n=308), Eurasian wren (n=61), Redwing (n=50), European robin (n=38), White wagtail (n=31), Broad-winged hawk (n=25) and European herring gull (n=18).
As expected, anthropogenic sounds were prevalent, with engine noise (n=73) and sirens (n=273) being the most frequently detected, followed by fireworks (n=175), likely associated with festivities during the deployment period.

The \textit{acoupi\_batdetect2} programme did not detect any bat echolocation calls with high confidence (score > 0.85), and the 174 pulses detected with moderate confidence (scores > 0.4) are likely false positives. 
This low number of bat detections is consistent with the deployment period (November), when most bat species in the UK are hibernating.
Detections were not validated post-deployment. See tables (\ref{tab:top20_detections_totalcount} and \ref{tab:bat_detection_summary} for a detailed summary of the detections made by the two bioacoustic classifiers.

\begin{table}[h]
\centering
\caption{Top 20 detection results for the \textit{acoupi\_birdnet} deployment sorted by total count.}
\renewcommand{\arraystretch}{1.5}
\adjustbox{width=\textwidth}{%
\fontsize{14}{18}\selectfont
\begin{tabular}{|>{\raggedright\arraybackslash}p{3.5cm}|>{\raggedright\arraybackslash}p{3.5cm}|>{\centering\arraybackslash}p{2.5cm}|>{\centering\arraybackslash}p{2.5cm}|>{\centering\arraybackslash}p{2cm}|>{\centering\arraybackslash}p{2.5cm}|>{\centering\arraybackslash}p{2.5cm}|>{\raggedright\arraybackslash}p{4cm}|}
\hline
\textbf{Latin Name} & \textbf{Common Name} & \textbf{Max Confidence Score} & \textbf{Mean Confidence Score} & \textbf{Total Count} & \textbf{Confidence Score > 0.4} & \textbf{Confidence Score > 0.85} & \textbf{Comment} \\ \hline
Engine & Engine & 0.9897 & 0.4050 & 7955 & 3151 & 73 & \\ \hline
Siren & Siren & 0.9994 & 0.5684 & 1382 & 911 & 273 & \\ \hline
Troglodytes troglodytes & Eurasian wren & 0.9772 & 0.5056 & 1308 & 844 & 61 & \\ \hline
Erithacus rubecula & European robin & 0.9949 & 0.4682 & 813 & 434 & 38 & \\ \hline
Fireworks & Fireworks & 0.9972 & 0.5722 & 782 & 491 & 175 & \\ \hline
Pica pica & Eurasian magpie & 0.9971 & 0.7191 & 707 & 613 & 308 & \\ \hline
Branta canadensis & Canada goose & 0.9508 & 0.3875 & 556 & 196 & 2 & \\ \hline
Larus argentatus & European herring gull & 0.9461 & 0.5207 & 326 & 215 & 18 & \\ \hline
Turdus iliacus & Redwing & 0.9974 & 0.5133 & 313 & 171 & 50 & \\ \hline
Turdus philomelos & Song thrush & 0.9461 & 0.4300 & 181 & 88 & 4 & \\ \hline
Buteo platypterus & Broad-winged hawk & 0.9946 & 0.5463 & 164 & 102 & 25 & False Detection. No such species in the UK \\ \hline
Cyanistes caeruleus & Eurasian blue tit & 0.9511 & 0.4075 & 108 & 48 & 2 & \\ \hline
Motacilla alba & White wagtail & 0.9959 & 0.6497 & 96 & 74 & 31 & \\ \hline
Anaxyrus microscaphus & Arizona toad & 0.9760 & 0.5097 & 78 & 47 & 10 & \\ \hline
Falco peregrinus & Peregrine falcon & 0.9973 & 0.5019 & 74 & 40 & 12 & \\ \hline
Strix aluco & Tawny owl & 0.9690 & 0.4264 & 64 & 27 & 4 & \\ \hline
Ardea cinerea & Grey heron & 0.9463 & 0.4262 & 46 & 15 & 2 & \\ \hline
Tyto alba & Barn owl & 0.9735 & 0.4656 & 36 & 23 & 1 & \\ \hline
Megaceryle alcyon & Belted kingfisher & 0.9751 & 0.5465 & 34 & 23 & 4 & False Detection. No such species in the UK \\ \hline
Cygnus cygnus & Whooper swan & 0.9597 & 0.4805 & 32 & 17 & 1 & \\ \hline
\end{tabular}%
}
\label{tab:top20_detections_totalcount}
\end{table}

\begin{table}[h]
\centering
\caption{Detection results for the \textit{acoupi\_batdetect2} deployment.}
\renewcommand{\arraystretch}{1.5}
\adjustbox{width=\textwidth}{%
\fontsize{14}{18}\selectfont
\begin{tabular}{|>{\raggedright\arraybackslash}p{4cm}|>{\raggedright\arraybackslash}p{4cm}|>{\centering\arraybackslash}p{2.5cm}|>{\centering\arraybackslash}p{2.5cm}|>{\centering\arraybackslash}p{4cm}|>{\centering\arraybackslash}p{2.5cm}|>{\centering\arraybackslash}p{2.5cm}|}
\hline
\textbf{Latin Name} & \textbf{Common Name} & \textbf{Max Confidence Score} & \textbf{Mean Confidence Score} & \textbf{Total Count} & \textbf{Confidence Score > 0.4} & \textbf{Confidence Score > 0.5} \\ \hline
Nyctalus leisleri & Lesser noctule & 0.6130 & 0.0175 & 1,604,336 & 170 & 0 \\ \hline
Plecotus austriacus & Grey long-eared bat & 0.4900 & 0.0125 & 205,073 & 1 & 0 \\ \hline
Pipistrellus nathusii & Nathusius's pipistrelle & 0.4650 & 0.0228 & 113,516 & 2 & 0 \\ \hline
Pipistrellus pipistrellus & Common pipistrelle & 0.4030 & 0.0317 & 658,435 & 1 & 0 \\ \hline
Myotis nattereri & Natterer's bat & 0.3870 & 0.0140 & 9,722 & 0 & 0 \\ \hline
Pipistrellus pygmaeus & Soprano pipistrelle & 0.3620 & 0.0316 & 208,803 & 0 & 0 \\ \hline
Nyctalus noctula & Common noctule & 0.3480 & 0.0223 & 240,926 & 0 & 0 \\ \hline
Plecotus auritus & Brown long-eared bat & 0.3030 & 0.0217 & 50,726 & 0 & 0 \\ \hline
Rhinolophus ferrumequinum & Greater horseshoe bat & 0.2860 & 0.0314 & 6,438,263 & 0 & 0 \\ \hline
Eptesicus serotinus & Serotine bat & 0.2700 & 0.0129 & 1,787,009 & 0 & 0 \\ \hline
Myotis alcathoe & Alcathoe bat & 0.2470 & 0.0195 & 38 & 0 & 0 \\ \hline
Rhinolophus hipposideros & Lesser horseshoe bat & 0.1970 & 0.0315 & 7,952,268 & 0 & 0 \\ \hline
Barbastellus barbastellus & Western barbastelle & 0.1810 & 0.0148 & 36,023 & 0 & 0 \\ \hline
Myotis bechsteinii & Bechstein's bat & 0.1380 & 0.0111 & 2,019 & 0 & 0 \\ \hline
Myotis daubentonii & Daubenton's bat & 0.0920 & 0.0113 & 97,757 & 0 & 0 \\ \hline
Myotis mystacinus & Whiskered bat & 0.0700 & 0.0096 & 8,344 & 0 & 0 \\ \hline
Myotis brandtii & Brandt's bat & 0.0220 & 0.0082 & 3,650 & 0 & 0 \\ \hline
\end{tabular}%
}
\label{tab:bat_detection_summary}
\end{table}

\end{document}